\documentclass[12pt]{article} \topmargin -5mm \oddsidemargin 7mm
\usepackage{graphicx,multirow}
\usepackage{amssymb,amsmath,amsfonts,bm}
\usepackage{colortbl}
\usepackage{natbib}

\definecolor{text1}{cmyk}{1,.65,0,0} 
\definecolor{text2}{cmyk}{0,.71,0.88,0.17} 
\definecolor{text2}{rgb}{1,0,0} 
\definecolor{text6}{rgb}{0,0.66,0} 
\definecolor{text3}{cmyk}{0,0,0,1} 
\definecolor{text4}{cmyk}{0,0,0,0.5} 

\title{\bf A Flexible Parametric Modelling Framework for Survival Analysis}
\author{Kevin Burke\\\emph{University of Limerick, Ireland}\\
M.C. Jones\\\emph{Open University, U.K.}\\
{Angela Noufaily}\\\emph{{University of Warwick, U.K.}}\\}
\date{}

\begin{document}

\maketitle

\bigskip

\begin{abstract}

We introduce a general, flexible, parametric survival modelling
framework
which encompasses key shapes of hazard function
(constant,
increasing, decreasing, up-then-down, down-then-up), various common survival
distributions (log-logistic, Burr type XII, Weibull, Gompertz), and
includes    defective distributions (i.e., cure models). This generality
is achieved using four basic distributional parameters: two scale-type
parameters
and two shape parameters. Generalising to covariate dependence,
the scale-type
regression components correspond to accelerated failure time (AFT) and
proportional hazards (PH) models. Therefore, this general formulation
unifies the most popular survival models which allows us to consider the
practical value of possible modelling choices for survival data.
Furthermore, in line with our proposed flexible baseline distribution, we
 advocate the use of multi-parameter regression in which more than one
 distributional parameter depends on covariates -- rather than the usual
 convention of having a single covariate-dependent (scale) parameter.
 While many choices are available, we suggest introducing covariates
 through just one or other of the two scale parameters, which covers
AFT
and PH models,
in combination with a ``power'' shape parameter, which allows for
more
complex non-AFT/non-PH effects,  while the {other} shape
parameter remains covariate-independent, and handles automatic
selection of the baseline distribution. We explore inferential
issues in simulations, both with and without a covariate, with particular
focus on evidence concerning the need, or otherwise, to include both AFT
and PH parameters.
We illustrate the
efficacy of our modelling framework by investigating differences
between treatment groups using data from a lung
cancer  study and a melanoma study. Censoring is accommodated throughout.
\end{abstract}

\section{Introduction\label{sec:intro}}

This article is concerned with both theoretical
and practical aspects of parametric survival
analysis with a view to providing an attractive and flexible general
modelling
approach to analysing survival data in areas such as medicine,
population health, and disease modelling.
In particular, focus will be {on} the choice of an appropriate
flexible form for
the distribution
of the survival outcome and the efficient use of multi-parameter
regression to understand the effects of covariates on survival.

We consider a
univariate lifetime random variable, $T>0$,
the primary survival outcome,
whose
cumulative hazard function (c.h.f.), $H(t)$, is, atypically
perhaps, modelled using a
flexible
parametric form which we take to be
\begin{equation}
 H(t) = \lambda \, H_0\left( \left(\phi t
\right)^\gamma;\kappa\right),~~~~~t>0.\label{main-framework}
\end{equation}
Here, $H_0(\cdot;\kappa)$ is an underlying c.h.f.\
with
shape parameter $\kappa$, and {$\phi>0$, $\lambda>0$ and $\gamma >0$}
are
further parameters with the following distinct interpretations:
 $\phi$ controls the
horizontal scaling of the hazard function, and is well known as
 the accelerated failure time parameter (also, $1/\phi$ is the usual
distributional scale parameter); $\lambda$ controls the vertical scaling of the hazard function, and is well
known as the proportional hazards parameter; and $\gamma$ is a second
shape
parameter which is explicitly defined as a power parameter (unlike $\kappa$ which can enter in potentially more complicated ways, and might even represent a vector of parameters). Were $Y = \log(T)$ to be modelled as a
location-scale distribution on $\mathbb{R}$, then $\mu = - \log \phi$ and
$\sigma = 1/\gamma$ would be the location and scale of that distribution,
respectively, these
relationships driving our preference to specify $\gamma$ as a
power
parameter rather than as a more general shape parameter. {As will be
clear in the   sequel, we intend that only one of the scale parameters be
present in the  model} {in order to avoid identifiability issues} {(i.e., we fix $\lambda = 1$ or $\phi = 1$).
However,
we write the model  in a general way for the purpose of unification of
sub-models.}

In this article, we also propose a specific choice for
$H_0(t^\gamma;\kappa)$, namely
\begin{equation}H_A({t;\gamma,\kappa}) = \frac{\kappa+1}{\kappa}
\left\{
\left(1+\frac{t^\gamma}{\kappa+1}\right)^{\kappa}-1\right\},
~~~~~t>0.
\label{adapted-chf}\end{equation}corresponding to an adapted form of the `power generalised Weibull'
(PGW) distribution
introduced
by \citet{bagnik:2002}; we will use APGW to stand for `adapted PGW'.
This choice
 has some major advantages: with just two shape parameters,  the
full range of simplest hazard shapes, namely, constant, increasing,
decreasing,
up-then-down or down-then-up (and no others), are available,
 the parameters $\gamma$ and $\kappa$ controlling this through the way
they control behaviour of the hazard function near zero and at infinity.
Here, we use the simple descriptive
terms
`up-then-down'
and `down-then-up' to avoid the term `bathtub-shaped', which is
down-then-up but with a flat valley, the clumsy term
`upside-down-bathtub-shaped', and the terms `unimodal/uniantimodal'
which also encompass monotone hazards.
Our adaptation of the PGW distribution also allows
$\kappa$ to control distributional
choice within the family:  for {$\kappa \geq 0$}, log-logistic and
Burr
Type XII
distributions are the
heaviest
tailed members, Weibull distributions {($\kappa=1$)} are `central'
within the
family, and
Gompertz-related
distributions are the most lightly tailed.
See Section \ref{sec:model}  for details of this model, which also include
its cure model special cases {when $-1<\kappa<0$}.

Any one or more of the four distributional parameters in model
(\ref{main-framework}) can be made to depend, typically log-linearly, on
covariates; such ``multi-parameter regression'' is one of the
focusses of this work. Indeed, this general formulation covers the most
popular survival models, e.g., the
accelerated
failure time (AFT) model
when $\phi$ depends on covariates,
 the proportional hazards (PH) model when
$\lambda$ depends on covariates, and semi-parametric versions when $H_0$
is an unspecified function. In particular, an advantage of
considering
(\ref{main-framework}) is that one may evaluate the breadth of
possible
modelling choices. Our primary focus in this respect is to consider which
distributional parameters should depend on covariates to assess, for
example, whether an AFT model ($\phi$ regression) is, in general, likely
to provide a superior fit when compared with a
PH
model ($\lambda$
regression), the utility of a simultaneous AFT-PH model (simultaneous
$\phi$ and $\lambda$ regression components;  \citet{chenjewell:2001})
{when $\kappa \neq 1$},
and the merits of a shape
regression component ($\gamma$ or $\kappa$) in addition to the, more
standard, AFT and PH components. One might also consider whether or not
non-parametric components should be introduced either for functions
of
covariates within the regression equations, or for the baseline
c.h.f., $H_0$,
or both. However, this is beyond the scope of the current paper.

The reason for our focus on the core model structure rather than the
development of non-/semi-parametric approaches is that, {within} the
survival literature, there is a general over-emphasis placed on
semi-parametric models -- compared with other fields of statistics -- to
the extent that many useful parametric alternatives do not receive the
attention they deserve. In particular, practitioners are often content
with the ``flexibility'' afforded by a non-parametric baseline function
without {concerning themselves with} the possibly inflexible structural assumptions
of the
model at hand. Indeed, a structurally flexible parametric framework has
the potential to outperform a less flexible semi-parametric model; for
example, there might be more to be gained by contemplating the
extension of a
PH model ($\lambda$ regression) to include a $\gamma$ {shape} regression, than
by extension to a non-parametric $H_0$. Of course, this is not to
downplay the importance of a sufficiently flexible baseline function, and
our proposed choice for $H_0$, (\ref{adapted-chf}), is
quite
general as it covers a wide variety of popular survival
distributions.

After Section \ref{sec:model}, in which we justify our
choice of baseline
distribution and develop its properties, we consider the extension
to regression modelling in Section~\ref{sec:regression}, including
model
interpretation and estimation. Then, the properties of estimation within
this general framework, and further practical aspects, are explored using
simulated and real data in Sections \ref{sec:sim} and
\ref{sec:real},
respectively. Finally, we close with some discussion in Section
\ref{sec:disc}.

\section{The Specific Model for {$H_0$}\label{sec:model}}

\subsection{Basic Definition and Properties}\label{sec:basics}

We recommend for general use the APGW distribution with c.h.f.\ given by
(\ref{adapted-chf}) and hazard function {is} \begin{equation}
h_A({t;\gamma,\kappa}) = \gamma t^{\gamma-1}
\left(1+\frac{t^\gamma}{\kappa+1}\right)^{\kappa-1},
~~~~~t>0.\label{adapted-hf} \end{equation} This is a tractable distribution with readily
available formulae for its (unimodal) density, survivor and quantile
functions also. It comes about by a particular vertical and horizontal
rescaling of the original PGW distribution which has c.h.f.
$H_N({t;\gamma,\kappa}) = (1+t^\gamma)^\kappa-1$ (see \citet{bagnik:2002},
\citet{nikhagh:2009} and \citet{dimetal:2007}; the $\gamma=1$ special case of $H_N$ is the
extended exponential distribution of \citet{nadarhagh:2011}).
 This resulting APGW distribution then retains attractive shape properties
of the PGW distribution's hazard function, includes important survival
distributions as special and limiting cases and extends to cure models, as
we now show.

First, for fixed
\(\gamma,\kappa>0\),
\[h_A({t;\gamma,\kappa}) \sim \gamma \, t^{\gamma-1}
\,\,\,\textrm{as}\,\,\,t \to
0~~~{\rm
and}~~~ h_A({t;\gamma,\kappa}) \sim (\kappa+1)^{1-\kappa}
\gamma\,
t^{\kappa\gamma-1}
\,\,\,\textrm{as}\,\,\,t \to \infty.\]
The power parameter \(\gamma\) controls the behaviour of the hazard
function at
zero: it  goes to $0 \,({\rm constant}) \,\infty$ as $ \gamma
>(=) < 1.$
As \(t \to \infty\), the hazard function goes to $ 0 \,({\rm constant})\,
\infty$ as $ \kappa \gamma < (=)\, > 1.$ In fact, the APGW hazard
function
joins these tails smoothly in such a way that its hazard shapes are
readily shown to be as
listed in Table~\ref{tab:pgwshapes}. Whenever the hazard function
is non-monotone, its mode/antimode is at
\(
\left\{(1-\gamma)(\kappa+1)/(\kappa\gamma-1)\right\}^{1/\gamma}.\)

\begin{table}[htbp]
\centering
\caption{Shapes of PGW hazard functions\label{tab:pgwshapes}}
\begin{tabular}{ccc}
\hline $\gamma$ & $\kappa\gamma$ & shape\\ \hline
1 & 1 & constant\\[-0.0cm]
$\leq 1$ & $\leq 1$ & decreasing\\[-0.0cm]
$\leq 1$ & $ \geq 1$ & down-then-up\\[-0.0cm]
$\geq 1$ & $\leq  1$ & up-then-down\\[-0.0cm]
$ \geq 1$ & $\geq 1$ & increasing\\[0.1cm]
 \hline
\multicolumn{3}{p{.3\textwidth}}{\footnotesize Here, pairs of $\leq$'s and/or $\geq$'s include the convention  `and not both equal at once'.}
\end{tabular}
\end{table}

Defining $H_A$ by (\ref{adapted-chf}) allows us to identify
an especially large number
of special and limiting cases, many important and well known, some
less
so, as listed in Table~\ref{tab:specialcases}.
(For  the `Weibull extension' distribution, see \citet{chen:2000} and
\citet{xieetal:2002}.) The shapes
of
their
hazard functions, which are also given in Table~\ref{tab:specialcases},
reflect the general shape properties of Table~\ref{tab:pgwshapes}, of course.

\begin{table}[htpb]
\centering
\caption{Special and limiting cases of APGW
distributions\label{tab:specialcases}}
\begin{tabular}{ccccc}
\hline $\kappa$ & $H_A$ &  shapes
of $h_A$& distribution  &  others encompassed \\
 \hline
&&&&\\[-0.3cm]
0 & $\log(1+t^\gamma)$ & decreasing, & log-logistic &  $H_A \times \lambda
\Rightarrow$ Burr
type XII \\[-0.0cm]
& & up-then-down &  \\[0.2cm]
1 & $t^\gamma$ &
decreasing, & Weibull &  $\gamma=1 \Rightarrow$
exponential \\[-0.0cm]
&& constant,&&\\[-0.0cm] & &increasing&& \\[0.2cm]
2 & $t^\gamma+\tfrac16 t^{2\gamma}$  & decreasing,  & &
$\gamma=1 \Rightarrow$ linear hazard \\[-0.0cm]
 & &down-then-up,  &   & \\[-0.0cm]&&increasing &&\\[0.2cm]
$\infty$& $e^{t^\gamma}-1$ & increasing,&  &
$H_A \times \lambda \Rightarrow$ Weibull extension; \\[-0.0cm]
&& down-then-up &  &   $H_A \times \lambda,\gamma = 1
\Rightarrow$  Gompertz \\[0.1cm] \hline
\end{tabular}
\end{table}

It can also be shown that the APGW distribution retains membership
of the
 log-location-scale-log-concave family of distributions of
\citet{jonesnouf:2015} and therefore,  inter alia, unimodality of
densities. We also now note, for future reference, the attractive
form of the quantile function associated with $H_A$, namely
$Q_A(u) = \{H_A{(-\log(1-u);1,1/\kappa)}\}^{1/\gamma}$ $\equiv
Q_{{A1}}(u;\kappa)^{1/\gamma}.$

The new adaptation can also be used to
widen
the family of PGW distributions by taking  \(-1<\kappa<0\). For clarity,
{temporarily} define $\psi = \kappa+1$ so that $0<\psi<1$. The APGW c.h.f.\
can then be written as
$$H_A({t;\gamma,\psi}) = \frac{\psi}{1-\psi} \left( 1-
\frac1{\{1+(t^\gamma/\psi)\}^{1-\psi}}\right).$$
This corresponds to a cure model with cure probability
$p_\psi \equiv {\lim_{t \to \infty}\exp(-H_A(t;\gamma,\psi))}$
$=\exp\{-\psi/(1-\psi)\}.$
Since the (improper) survival function is in this case of the form
$p_\psi^{1-{S_C}(t)}$,
this cure model has  an interpretation as the distribution of the minimum
of a Poisson number of random variables (e.g.\ cancer cells,
tumours), each following the lifetime distribution with survival function
{$S_C$} (e.g.\ \citet{tsodikovetal:2003}); here, the Poisson parameter
is
$\psi/(1-\psi)$ and ${S_C}(t)=
\{1+(t^\gamma/\psi)\}^{\psi-1}$ is the survival
function of a scaled Burr Type  XII distribution.
The hazard  functions {$h_A(t;\gamma,\psi)$} follow the
shape of  their $\psi \to 1$ limit --- the log-logistic --- being
decreasing  for $\gamma \leq 1$
and up-then-down otherwise.

The PGW distribution, and in slightly more complicated form the APGW
distribution, exhibit interesting frailty relationships between
members of the families with different values of $\kappa$. We defer
consideration of these frailty links to  \citet{jnb:2018} where we exploit them to obtain a useful
bivariate shared frailty model with PGW/APGW marginal distributions.
In addition, PGW and APGW distributions are written as linear
transformation models in
the Appendix.

\subsection{Why This Particular Choice for $H_0$?}

\begin{figure}[!htbp]
\begin{center}
\includegraphics[width=0.9\textwidth, trim = {1.25cm 0cm 0cm 0cm}
]{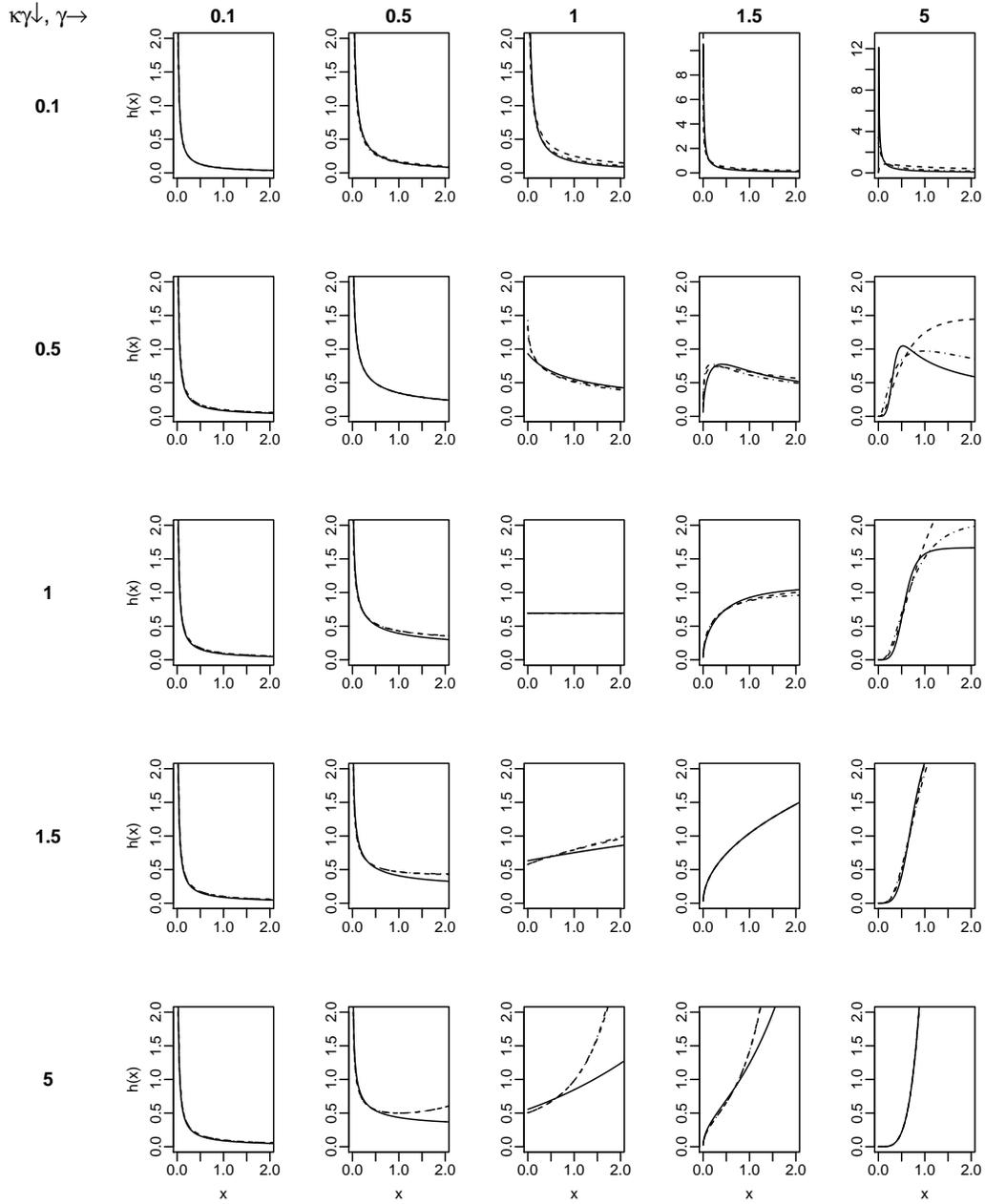}
\caption{Hazard functions of PGW
(solid), GG (dashed), and EW
(dot-dashed) distributions for the values of $\gamma$ and
$\kappa\gamma$ specified along the top and down the left-hand side of
the figure, respectively. In each case, the scale parameter is chosen such
that the
median of the distributions is one. Figures on the main diagonal of the
matrix of figures, in each of which the PGW, GG and EW hazard
functions are
identical, correspond to Weibull
distributions, the figure in the centre to the exponential
distribution.\label{fig:comparedist}} \end{center}
\end{figure}

The PGW/APGW distribution shares the set of hazard behaviours listed
in Table~\ref{tab:pgwshapes} with
two other established two-shape-parameter lifetime distributions
centred on the Weibull distribution, namely, the
generalised  gamma (GG) and exponentiated Weibull (EW) distributions; see
\citet{jonesnouf:2015}.
See Figure~\ref{fig:comparedist} for many examples of just how similar the hazard shapes of all
three distributions are; in Figure~\ref{fig:comparedist},  we have chosen
the scale parameter
such that each distribution has median one, used the PGW vertical
scaling and otherwise
specified shape parameters $\gamma,\kappa>0$ only so that all three
hazard functions behave as $t^{\gamma-1}$ as $t \to 0$ and as
$t^{\kappa\gamma-1}$ as $t \to \infty$.

Further effort to choose shape
parameters to match hazard functions or other aspects of the distributions
even more closely is possible and has been pursued for the EW and GG
distributions by \citet{coxmath:2014} and extended to the PGW distribution
(what they call the generalised Weibull distribution) by
\citet{mathetal:2017}. \citet{coxmath:2014} state that the ``agreement
between the two distributions [GG and EW] in our various comparisons, both
graphically and in
terms of the K--L [Kullback-Leibler] distance, is striking'';
after a similar K--L matching exercise, \citet{mathetal:2017} state that
``the survival and hazard functions of the [PGW]
distribution
and its matched GG are visually indistinguishable.'' It remains, therefore,
to choose between APGW, GG and EW distributions on other grounds.
The GG
distribution includes the Weibull and gamma distributions as special cases
and the lognormal as a limiting one; the EW distribution includes the
Weibull and exponentiated exponential distributions.
However: we have been unable to match the number of
APGW's advantageous properties as in the previous subsections by
similarly adapting either the   GG or
EW distributions; we prefer the breadth of/difference between the
wide range of distributions encompassed by the APGW distribution;
and
we appreciate the
greater tractability of
 the APGW distribution both mathematically and computationally
(for
instance,  its hazard
function has a simpler form
compared with the GG --- which involves an incomplete gamma function --- and the
EW).

\section{Regression\label{sec:regression}}

\subsection{Modelling Choices\label{sec:modchoices}}

Within our proposed APGW modelling framework, there are four parameters,
$\phi, \lambda,\, \gamma$, and $\kappa$, which could potentially
depend on
covariates. Note that most classical modelling approaches are based on
having a \emph{single} covariate-dependent distributional parameter, which
we refer to as single parameter regression (SPR), where,
understandably,
there is a particular emphasis on scale-type parameters, e.g.,  the
accelerated failure time (AFT) model ($\phi$ regression) and  the
proportional hazards (PH) model ($\lambda$ regression).
However,
 in
 line with the flexibility of the APGW distribution itself, we also
 consider taking a flexible multi-parameter regression (MPR) approach in
 which more than one parameter may depend on covariates
(cf.~\citet{burmac:2017}, and references therein, for details of
multi-parameter regression). The most general linear APGW-MPR is,
therefore, given by
\begin{align*}
\log(\phi) &= x^T \tau, & \log(\lambda) &= x^T \beta, &
\log(\gamma) &= x^T \alpha, & \log(\kappa+1) &= x^T \nu,
\end{align*}
where log-link functions are used to respect the positivity of the
parameters $\phi$, $\lambda$ and $\gamma$, with a slightly
different link
function for $\kappa$ to accommodate the fact that, within our APGW,
it can take values in the range $(-1, \infty)$ (see Section
\ref{sec:basics}), $x = (1, x_1, \ldots, x_p)^T$ is a vector of
covariates,
and $\tau = (\tau_0,
\tau_1, \ldots, \tau_p)^T$, $\beta = (\beta_0, \beta_1, \ldots,
\beta_p)^T$,  $\alpha = (\alpha_0, \alpha_1,$ $\ldots,
\alpha_p)^T$, and $\nu = (\nu_0, \nu_1, \ldots, \nu_p)^T$ are the
corresponding vectors of regression coefficients. In practice,  we may not
necessarily have the same set of covariates appearing in all   regression
components, and, in our current notation, this can be handled by  setting
various regression coefficients to zero.

As mentioned in Section \ref{sec:intro}, we could extend the above
regression specification via non-parametric regression functions of
$x$, but this is beyond the scope of this paper and, indeed, the MPR
approach is, in itself, already flexible without this added
complexity. Furthermore, although the general APGW-MPR model offers the
opportunity of four regression components simultaneously, this full
flexibility is unlikely to be required in practice. {In particular, it
is well known} {that} {$\phi$ and $\lambda$ coincide in the
Weibull distribution} {so}
{that only one scale parameter is needed in this case,
i.e., $\phi$ and $\lambda$ are} {non-identifiable
when $\kappa=1$. Moreover}{,
our numerical studies (Sections \ref{sec:sim}  and \ref{sec:real}) suggest
that, outside of the Weibull, this is} {effectively} {much more
generally true.
Specifically, although $\phi$ and $\lambda$ are  theoretically
identifiable in the non-Weibull cases, the parameters  are nonetheless
nearly non-identifiable in finite samples, which is an  apparently new
finding in the literature. Thus, in general, we should  fix either
$\phi=1$ or $\lambda=1$, but we would not simultaneously  fix
$\phi=\lambda=1$ as a scale parameter is a core component  for most
statistical models.}

A good practical choice is composed of the
following pieces: (a) only one scale parameter ($\phi$ or $\lambda$)
depends on covariates, {while the other is fixed at one as mentioned above}, (b) the
$\gamma$ shape parameter may depend on covariates, and (c) the $\kappa$
shape parameter is constant, i.e., only the intercept, $\nu_0$, is
non-zero in the $\nu$ vector. This choice provides a useful framework
which incorporates, depending on the choice of scale regression, either
an
AFT ($\tau$) or PH ($\beta$)  component, allows for non-AFT/non-PH
effects
via the $\alpha$ coefficients associated with the power parameter
(Section \ref{sec:interpretation}), and
automatically selects the underlying baseline distribution via $\nu_0$
from a range of popular survival distributions (Table
\ref{tab:specialcases}) including defective distributions, i.e., cure
models (Section \ref{sec:basics}).


\subsection{{Suggested models: \boldmath $M(\tau, \alpha)$ and $M(\beta, \alpha)$}\label{sec:interpretation}}

 Let $M(\tau,\alpha)$ and $M(\beta,\alpha)$ denote the
two
models
 {suggested} in the previous paragraph, e.g., the latter is the
model
with $\beta$ and
 $\alpha$ regression components, along with the shape parameter $\nu_0$
 (but where $\tau$ is a vector of zeros). More generally, beyond these two
  models, we will use this notation throughout the paper where
 the arguments of $M(\cdot)$ indicate which regression components are
 present in the model, the absence of either $\beta$ or $\tau$
indicating
 that this is a vector of zeros. Irrespective of the presence of
 $\alpha$ or $\nu$ regression components {in this $M(\cdot)$ notation}, we assume that $\alpha_0$ and
 $\nu_0$ are always present since these are needed to characterise the
 baseline distribution and the shape of its hazard function (see Tables
 \ref{tab:pgwshapes} and \ref{tab:specialcases}). Thus, for example,
  $M(\tau)$ and $M(\beta)$ are, respectively, AFT and PH
models with two
 shape parameters ($\alpha_0$ and $\nu_0$), $M(\beta, \alpha, \nu)$ is a
 model which extends the suggested  $M(\beta, \alpha)$ model so that the
 $\nu$ regression component is {also}  present, and $M(\beta, \tau,
\alpha, \nu)$
 is the most general APGW-MPR model.

We first consider model $M(\tau, \alpha)$ which extends the basic AFT
model, {$M(\tau)$}, via the incorporation of the $\alpha$ regression component. Now
suppose that $x_j$ is a binary covariate and let
$x_{(-j)} = (1,x_1,\ldots,x_{j-1},0,$ $x_{j+1},\ldots,x_p)^T$ be the
covariate vector with $x_j$ set to zero so that we may write
$x^T \tau = x_j \tau_j + x_{(-j)}^T \tau$ and $x^T \alpha = x_j \alpha_j +
x_{(-j)}^T \alpha$. As this model extends the AFT model, it is
natural to consider its quantile function which is given by
\begin{align*} Q(u|x) = \exp(-x^T\tau) \,
Q_{{A1}}(u;\kappa)^{\exp(-x^T\alpha)}
\end{align*}
where $Q_{{A1}}(u) = H_A{(-\log(1-u);1,1/\kappa)}$ is the
``baseline''
quantile
function defined {in} Section~\ref{sec:basics}.  We can
then inspect the quantile ratio
\begin{align*}
QR_j(u) = \frac{Q(u|x_j=1)}{Q(u|x_j=0)} = \exp(-\tau_j)
\,Q_{{A1}}(u;\kappa)^{\exp(- x_{(-j)}^T \alpha)\{\exp(-\alpha_j)-1\}}
\end{align*}
where we see that $\alpha_j$ is the  key parameter in determining the
$u$-dependence. In particular, since  $Q_{\rm A1}(u;\kappa)$ is an
increasing
function of $u$, $QR_j(u)$ increases when  $\alpha_j < 0$, decreases
when
$\alpha_j > 0$, and is constant (i.e., the usual AFT case) when
$\alpha_j =
0$. Hence, the $\alpha_j$ coefficient {characterises} the nature of
the  effect of the binary covariate
$x_j$, and
provides a test of the AFT property for that covariate.

Now consider the model $M(\beta, \alpha)$ which extends the PH model, {$M(\beta)$} , and
whose hazard function is given by
\begin{align*} h(t|x) = \exp(x^T\beta)\,
h_A({t;{\exp(x^T\alpha)},\kappa})
\end{align*}
where $h_A({t;\gamma,\kappa})$
is defined in (\ref{adapted-hf}).  The hazard ratio for the binary
covariate $x_j$ is
\begin{align*} HR_j(t) = \frac{h(t|x_j=1)}{h(t|x_j=0)}
= \exp(\beta_j+\alpha_j) \, t^{\exp(x_{(-j)}^T
\alpha)\{\exp(\alpha_j)-1\}} g(t;\alpha_j, x_{(-j)}^T\alpha,\kappa)
\end{align*}
where
\begin{align*}g(t;\alpha_j, x_{(-j)}^T\alpha,\kappa)=
\left(\frac{t^{\exp(\alpha_j + x_{(-j)}^T \alpha)}+\kappa +
1}{t^{\exp(x_{(-j)}^T \alpha)}+\kappa + 1}\right)^{\kappa-1} .
\end{align*}
Clearly, $\alpha_j$ characterises departures from
proportional hazards as $HR_j(t)$ is a constant when $\alpha_j = 0$.
For $\kappa \ge 0$, we have that $\lim_{t\to0}HR_j(t) = 0$ and $\lim_{t\to\infty}HR_j(t) = \infty$ when $\alpha_j>0$, while $\lim_{t\to0}HR_j(t) = \infty$ and $\lim_{t\to\infty}HR_j(t) = 0$ when $\alpha_j<0$. Furthermore, it can be shown that $HR_j(t)$ varies monotonically in $t$ in the following cases: (i) $\kappa \geq 1$, or (ii) $0 < \kappa < 1$ and
$\alpha_j \notin (\log\kappa,-\log\kappa)$. (We do not know about
monotonicity or otherwise in the remaining cases.)

 From the above we see that, within the suggested $M(\tau, \alpha)$ and
$M(\beta, \alpha)$ models, the parameters play {the following} roles: the scale
coefficients ($\tau$ or $\beta$) control the overall size of the effect
where negative coefficients correspond to longer lifetimes; the
$\alpha$
shape coefficients describe how covariate effects vary over the lifetime,
i.e., permitting non-AFT and non-PH effects; and the $\nu_0$ shape
parameter characterises the baseline distribution. Note that we could,
alternatively, achieve non-constant {covariate} effects via the $\nu$ regression
component rather than the $\alpha$ component, i.e., using
$M(\beta,\nu)$ rather than $M(\beta,\alpha)$. However, in this case, the
interpretation is that such non-constant effects are due to populations
which arise from structurally different distributions, rather than
different shapes within a given baseline distribution. The latter is
arguably more natural as it creates a separation of parameters
whereas, in the former, distribution selection and non-constant {covariate} effects
are intertwined. Of course, this is not to say that models with $\nu$
components instead of, or in combination with, $\alpha$ components will
never be useful in practice. However, we are highlighting practical merits
of the $M(\tau, \alpha)$ and $M(\beta, \alpha)$ models and, indeed, the
general use of these models is motivated by the numerical studies of
Sections \ref{sec:sim} and \ref{sec:real}.

\subsection{Estimation}

Consider the model formulation given in (\ref{main-framework}) with   all
four regression components, i.e., the $M(\tau, \beta, \alpha, \nu)$ model.
{(While we advocate the special cases $M(\tau, \alpha)$ or $M(\beta,
\alpha)$, we  write the estimation equations in a general form below so as
to unify all potential model structures.} {In particular, estimation
of both $\tau$ and $\beta$ is not recommended in practice.)} Let  $\phi_i
=
\exp(x_i^T \tau)$, $\lambda_i = \exp(x_i^T
\beta)$,
$\gamma_i =
\exp(x_i^T \alpha)$ and $\kappa_i = \exp(x_i^T \nu) - 1$ be the
covariate-dependent distributional  parameters for the $i$th individual
with covariate vector $x_i = (1,  x_{i1}, \ldots, x_{ip})^T$, and
$\tau$, $\beta$,
$\alpha$, and $\nu$ are  the associated vectors of regression
coefficients. Allow independent censoring by attaching to each individual an
indicator $\delta_i$ which equals one if the response is observed, and zero if it is
right-censored. The log-likelihood function is then given by
\begin{align*}
\ell(\theta) &= \sum_{i=1}^n \left[\delta_i \left\{\log\left(\frac{\lambda_i \gamma_i z_i}{t_i}\right) + m_0(z_i;\kappa_i)\right\} - \lambda_i H_0(z_i;\kappa_i)\right]
\end{align*}
where $\theta = (\tau^T, \beta^T, \alpha^T, \nu^T)^T$, $z_i =
(\phi_i t_i)^{\gamma_i}$ and,
in our proposed APGW case,
$$
H_0({t;\kappa}) = {H_A(t;1,\kappa) =}\,
\frac{\kappa+1}{\kappa}\left\{\left(1+\frac{t}{\kappa+1}\right)^\kappa-1\right\},$$
$$m_0(t;\kappa) = \log h_0(t;\kappa) = (\kappa-1) \log
\left(1+\frac{t}{\kappa+1}\right).$$
As usual, the log-likelihood function can be maximised by  solving the
score equations
\begin{align*}
(U_\tau^T X,~U_\beta^T X,~U_\alpha^T X,~U_\nu^T X)^T = 0_{4p \times
1}
\end{align*}
where $X$ is an $n\times p$ matrix whose $i$th row is $x_i$, $0_{4p \times
1}$ is a $4p \times 1$ vector of   zeros and  $U_\tau$, $U_\beta$,
$U_\alpha$, and $U_\nu$ are $n\times 1$ vectors whose $i$th elements are
as follows:
\begin{align*}
U_{\tau,i}
&= \delta_i \left\{\gamma_i + \gamma_i z_i \,m_0'(z_i;\kappa_i)  \right\}
- \lambda_i\gamma_i z_i\, h_0(z_i;\kappa_i), \\[0.3cm]
U_{\beta,i}
&= \delta_i - \lambda_i H_0(z_i;\kappa_i), \\[0.3cm]
U_{\alpha,i}
&= \delta_i\left[1 +  \log(z_i)\,\left\{1 + z_i\,
m_0'(z_i;\kappa_i)
\right\} \right] - \lambda_i z_i \log(z_i) h_0(z_i;\kappa_i), \\[0.3cm]
U_{\nu,i}
&= \left[ \delta_i  \left\{\frac{\kappa_i+1}{\kappa_i-1} m_0(z_i;\kappa_i)  - z_i m_0'(z_i;\kappa_i)\right\} \right. \\
 & \left. \qquad\quad- \lambda_i  \frac{\kappa_i+1}{\kappa_i^2{(\kappa_i-1)}}
\left\{{\kappa_i - 1 +
\left(1+\frac{t}{\kappa_i+1}\right)^{\kappa_i}}
a_0(z_i;\kappa_i)\right\}\right],
\end{align*}
where $a_0(t;\kappa) = \kappa(\kappa+1)m_0(t;\kappa)- \kappa^2
 t\,m_0'(t;\kappa) - \kappa + 1$. 

Note that the vectors $U_{\tau,i}$,  $U_{\beta,i}$ and
$U_{\alpha,i}$ are
written generically so that they apply to any model of the form given in
(\ref{main-framework}), i.e., they are not specific to the APGW case;
the form of  $U_{\nu,i}$, on  the other hand, uses the way in which $H_0$
and hence $m_0$ and {$m_0'$}
depend on $\kappa.$  Thus,
although the APGW is certainly a flexible choice (see Section
\ref{sec:model}), the first three score components extend
immediately to other
baseline distributions by replacing $H_0$ (and, consequently, $m_0$ and
$m_0'$). Estimation then proceeds {once} the functional form of $U_{\nu,i}$ has been
re-evaluated.

Furthermore, one may, alternatively, prefer to
maintain an unspecified baseline distribution, whereby $\nu$ represents an
infinite-dimensional (possibly covariate-\linebreak independent) parameter
vector. In this
case, estimation equations for the regression coefficients $\tau$,
$\beta$, and $\alpha$ can be based on $(U_\tau^T X,~U_\beta^T
X,~U_\alpha^T
X)$ where $H_0$ is replaced with {an appropriate non-parametric}
estimator (and, similarly, for $m_0$ and $m_0'$).  However, while
non-parametric estimation of $H_0$ is {reasonably} straightforward  (say, using a
Nelson-Aalen-type estimator), it is well known that  terms such as $m_0$
and $m_0'$, which involve $h_0$ and $h'_0$, are more difficult to
estimate consistently.  We note that
semi-parametric versions of the $M(\tau,\beta)$ and $M(\tau,\alpha)$
models have respectively been developed by \citet{chenjewell:2001} and
\citet{burerikpip:2018}. However, we are unaware of a semi-parametric
$M(\tau,\beta,\alpha)$ model in the literature. In any case, such
semi-parametric models are beyond the scope of the current paper and,
indeed, a flexible parametric framework can cover a wide variety of
applications as previously discussed in Section \ref{sec:intro}.

\section{Simulation Studies\label{sec:sim}}

\subsection{Without Covariates\label{sec:nocov}}

{Before considering estimation in the presence of covariates, we first investigate estimation in the context of the APGW model with no covariates. Thus, } we simulated  data from the APGW distribution parameterised  in terms of the following unconstrained parameters:  $\tau =
\log \phi$,
$\beta = \log \lambda$,  $\gamma = \log \alpha$ and $\nu = \log (\kappa+1)$. The
values of the first three parameters were fixed at $\tau = 0.8$, $\beta =
0.5$, $\alpha = -0.3$, respectively, while $\nu$ was varied such that $\nu
\in \{0.00, 0.22, 0.41, 0.69, 1.10, 1.61,  \infty\}$ (rounded to two
decimal places); note that $\nu = 0$, $\nu = 0.69$ and $\nu = \infty$
correspond, respectively, to the log-logistic ($\kappa=0$), Weibull ($\kappa=1$) and Gompertz ($\kappa=\infty$)
distributions. Furthermore, the sample size was fixed at 1000 and
censoring times were generated from an exponential distribution such that,
for each $\nu$ value, the censoring rate was fixed at approximately 30\%.
Within each of the seven simulation scenarios (i.e., varying $\nu$), we
fitted {three} different models with the aim of understanding the {identifiability of parameters in a finite, but reasonably large, sample}: (i) estimate all parameters,
(ii) fix $\beta$ at its true value, {and} (iii) fix $\beta$ at zero.
  Thus, $\tau${, $\alpha$ and $\nu$} are estimated in
all {three} models.
We also considered additional scenarios where
$\alpha = 0.3$ but the results are similar and, therefore, are not shown here.

Each scenario was replicated 1000 times,  and the results are contained in
 Table~\ref{tab:nocov}. {Clearly, estimation is unstable in model (i),
 i.e., standard errors are large.} {This} {instability arises
as a
 consequence of attempting to estimate the scale parameters, $\tau$ and
 $\beta$, simultaneously.} Indeed, in all cases where these parameters are
estimated
simultaneously, we have found that corr$(\hat\tau,\hat\beta) \approx 1$.
Of course, {it is well known that} $\hat\tau$ and $\hat\beta$ are perfectly co-linear in the
Weibull case ($\nu = 0.69$), but it is interesting to find that this
extends (approximately) beyond the Weibull distribution. This appears to
be a new finding in survival modelling and implies that these parameters
play {somewhat} similar roles across a range of popular lifetime
distributions {(it also   explains the large standard errors observed}
{in}
{Table 2 of \citet{chenjewell:2001}). This instability vanishes once
$\beta$ is fixed. In particular,} {when $\beta$ is set to
its true value of} {$\beta=0.5$  (i.e., model (ii)),  the
estimates display very little bias.} {Moreover,  when $\beta$ is set
to an incorrect value,} {$\beta=0.0$ (i.e.,
model (iii)),} $\hat\tau$ converges {consistently} to a value in
the range 1.4--1.5 which {compensates for the incorrect
specification of $\beta$ and} varies smoothly
with $\nu$; {the value of
$\hat\nu$ changes somewhat from its value in model (ii), but, interestingly, $\hat\alpha$ does not.}
Furthermore, the fitted  survivor curves for both models (not shown) are
close to the truth, i.e., there is no
reduction in quality of  model fit as a consequence of
fixing $\beta$ to {zero.  Similarly (but not shown here), estimation
is also stable if $\tau$ is  fixed and $\beta$ is estimated, and the
fitted survivor curves are again  close to the truth. Therefore, the
choice of scale --- either $\tau$  or $\beta$ (fixing the other to zero)
--- behaves, approximately, as a  model reparameterisation (which it is,
exactly, in the Weibull case).} {We note that, for both models (ii)
and  (iii),} the standard {error} of $\hat\nu$ can be large when $\nu$
is large.  {However,} this is not a concern as it is a consequence of
the fact that the APGW distribution changes very little over a range of large $\nu$ values.

\begin{table}[htbp]
\begin{center}
\caption{Median and standard {error} (in brackets) of estimates\label{tab:nocov}}
\begin{footnotesize}
\begin{tabular}{r@{~~}l@{\qquad}c@{\qquad}r@{~}c@{\quad}r@{~}c@{\quad}r@{~}c@{\quad}r@{~}c}
\hline
&&&&&&&&&&\\[-0.4cm]
\multicolumn{2}{c}{Model}      & $\nu$ & \multicolumn{2}{c}{$\hat\tau$} & \multicolumn{2}{c}{$\hat\beta$} & \multicolumn{2}{c}{$\hat\alpha$} & \multicolumn{2}{c}{$\hat\nu$} \\
\hline
&&&&&&&&&&\\[-0.3cm]
\multirow{7}{*}{(i)} & \multirow{6}{*}{$\beta$~:~est}
                                &     0.00  &      1.87  &   (6.93)   &  -0.21  &    (4.98)  &   -0.26  &   (0.12)  &     0.18  &   (4.38) \\[-0.0cm]
&\multirow{6}{*}{$\nu$~:~est}   &     0.22  &      2.24  &   (11.52)  &  -0.54  &    (8.38)  &   -0.26  &   (0.15)  &     0.39  &   (7.24) \\[-0.0cm]
                               &&     0.41  &      2.83  &   (11.55)  &  -0.98  &    (8.58)  &   -0.26  &   (0.20)  &     0.49  &   (7.55) \\[-0.0cm]
                               &&     0.69  &      2.11  &   (8.52)   &  -0.48  &    (6.54)  &   -0.31  &   (0.22)  &     0.76  &   (7.93) \\[-0.0cm]
                               &&     1.10  &      0.83  &   (4.30)   &   0.38  &    (3.32)  &   -0.33  &   (0.14)  &     1.18  &   (6.82) \\[-0.0cm]
                               &&     1.61  &      0.53  &   (2.76)   &   0.57  &    (2.13)  &   -0.32  &   (0.11)  &    12.32  &   (6.47) \\[-0.0cm]
                               &&  $\infty$ &      1.10  &   (1.53)   &   0.19  &    (1.21)  &   -0.32  &   (0.09)  &    13.04  &   (6.40) \\[0.1cm]
\hline
&&&&&&&&&&\\[-0.3cm]
\multirow{7}{*}{(ii)} & \multirow{6}{*}{$\beta$~:~true}
                                &     0.00  &      0.81  &   (0.15)   &   0.50  &     ---    &    -0.30  &   (0.05)  &     0.00  &   (0.11) \\[-0.0cm]
&\multirow{6}{*}{$\nu$~:~est}   &     0.22  &      0.79  &   (0.15)   &   0.50  &     ---    &    -0.30  &   (0.05)  &     0.23  &   (0.15) \\[-0.0cm]
                               &&     0.41  &      0.80  &   (0.15)   &   0.50  &     ---    &    -0.30  &   (0.05)  &     0.40  &   (0.19) \\[-0.0cm]
                               &&     0.69  &      0.79  &   (0.15)   &   0.50  &     ---    &    -0.30  &   (0.06)  &     0.71  &   (0.31) \\[-0.0cm]
                               &&     1.10  &      0.79  &   (0.16)   &   0.50  &     ---    &    -0.30  &   (0.06)  &     1.12  &   (1.59) \\[-0.0cm]
                               &&     1.61  &      0.80  &   (0.14)   &   0.50  &     ---    &    -0.30  &   (0.05)  &     1.65  &   (3.72) \\[-0.0cm]
                               &&  $\infty$ &      0.84  &   (0.09)   &   0.50  &     ---    &    -0.28  &   (0.04)  &    13.05  &   (6.23) \\[0.1cm]
\hline
&&&&&&&&&&\\[-0.3cm]
\multirow{7}{*}{(iii)} & \multirow{6}{*}{$\beta$~:~zero}
                                &     0.00  &      1.52  &   (0.12)   &   0.00  &     ---    &   -0.29  &   (0.05)  &     0.15  &   (0.09) \\[-0.0cm]
&\multirow{6}{*}{$\nu$~:~est}   &     0.22  &      1.50  &   (0.13)   &   0.00  &     ---    &   -0.29  &   (0.06)  &     0.33  &   (0.12) \\[-0.0cm]
                               &&     0.41  &      1.49  &   (0.13)   &   0.00  &     ---    &   -0.30  &   (0.05)  &     0.48  &   (0.14) \\[-0.0cm]
                               &&     0.69  &      1.48  &   (0.13)   &   0.00  &     ---    &   -0.30  &   (0.06)  &     0.68  &   (0.19) \\[-0.0cm]
                               &&     1.10  &      1.44  &   (0.13)   &   0.00  &     ---    &   -0.31  &   (0.06)  &     0.99  &   (0.27) \\[-0.0cm]
                               &&     1.61  &      1.44  &   (0.14)   &   0.00  &     ---    &   -0.31  &   (0.06)  &     1.27  &   (1.46) \\[-0.0cm]
                               &&  $\infty$ &      1.42  &   (0.13)   &   0.00  &     ---    &   -0.31  &   (0.06)  &     2.04  &   (4.50) \\[0.1cm]
\hline
&&&&&&&&&&\\[-0.3cm]
\multicolumn{11}{p{.8\textwidth}}{\scriptsize All numbers are rounded to two decimal places. For the models with fixed parameters, the ``estimated'' value shown is the value at which the parameter is fixed, and its standard error is then indicated by ``---''. While $\tau$ and $\beta$ are not simultaneously estimable when $\nu = 0.69$ (Weibull case), the estimation procedure still yields values (which depend completely on initial values) such that the constant $\lambda \phi^\gamma$ is preserved in the sense that its value is the same as the case where one of $\beta$ or $\tau$ were held constant.}
\end{tabular}
\end{footnotesize}
\end{center}
\end{table}

\subsection{With a Covariate\label{sec:simcov}}

We simulated survival times according to the APGW distribution with
parameters $\phi =  \exp(\tau_0 + \tau_1 X)$, $\lambda = \exp(\beta)$,
$\gamma = \exp(\alpha_0 + \alpha_1 X)$, and $\kappa = \exp(\nu)-1$ where
$X \sim \text{Bernoulli}(0.5)$, $\nu$ was varied according to the set
$\{0.00, 0.22, 0.41, 0.69, 1.10, 1.61, \infty\}$, and the remaining
parameter values were fixed at $\tau_0 = 0.8$, $\tau_1 = 0.6$, $\beta =
0.0$, $\alpha_0 = 0.2$, and $\alpha_1 = -0.5$; these values were selected
to yield realistic survival times. In the notation of Section
\ref{sec:interpretation},  the true model is  $M(\tau,\alpha)$. As in Section \ref{sec:nocov}, the sample size and censored proportion were, respectively, set at 1000 and 30\% (with censoring times generated from an exponential distribution). Within each of the seven scenarios (i.e., varying $\nu$), we fitted the following three regression models: the more general $M(\tau,\beta,\alpha)$, the true $M(\tau,\alpha)$, and the misspecified $M(\beta,\alpha)$, respectively. 
The
results, based on 1000 simulation replicates,  are given
in Table \ref{tab:cov}.

\begin{table}[htbp]
\begin{center}
\caption{Median and standard {error} (in brackets) of estimates\label{tab:cov}}
\begin{footnotesize}
\begin{tabular}{c@{\qquad}r@{~}c@{~~}r@{~}c@{~~}r@{~}c@{~~}r@{~}c@{~~}r@{~}c@{~~}r@{~}c@{~~}r@{~}c}
\hline
&&&&&&&&&&&&&&\\[-0.3cm]
& \multicolumn{14}{c}{\underline{Model$(\tau, \beta, \alpha)$}} \\[-0.0cm]
&&&&&&&&&&&&&&\\[-0.3cm]
$\nu$ &  \multicolumn{2}{c}{$\hat\tau_0$} & \multicolumn{2}{c}{$\hat\tau_1$} & \multicolumn{2}{c}{$\hat\beta_0$} & \multicolumn{2}{c}{$\hat\beta_1$} & \multicolumn{2}{c}{$\hat\alpha_0$} &  \multicolumn{2}{c}{$\hat\alpha_1$} & \multicolumn{2}{c}{$\hat\nu_0$} \\[-0.0cm]
 \hline
&&&&&&&&&&&&&&\\[-0.3cm]
   0.00  &   0.98 & (1.21) &   0.55 & (1.28) &   -0.20 & (1.37) &   0.02 &  (0.70) & 0.23 & (0.13) &  -0.50 & (0.18) &     0.05 & (0.76) \\[-0.0cm]
   0.22  &   0.98 & (1.40) &   0.51 & (2.08) &   -0.20 & (1.63) &   0.04 &  (1.61) & 0.23 & (0.17) &  -0.50 & (0.24) &     0.27 & (0.44) \\[-0.0cm]
   0.41  &   1.00 & (1.65) &   0.57 & (2.95) &   -0.24 & (1.99) &   0.10 &  (2.48) & 0.24 & (0.18) &  -0.51 & (0.26) &     0.40 & (0.39) \\[-0.0cm]
   0.69  &   1.18 & (2.13) &   0.34 & (4.39) &   -0.43 & (2.63) &   0.19 &  (3.85) & 0.22 & (0.20) &  -0.51 & (0.22) &     0.61 & (3.56) \\[-0.0cm]
   1.10  &   0.39 & (1.50) &   0.08 & (2.88) &    0.35 & (1.89) &   0.21 &  (2.66) & 0.17 & (0.12) &  -0.49 & (0.18) &     1.31 & (6.88) \\[-0.0cm]
   1.61  &   0.55 & (1.00) &   0.47 & (1.62) &    0.26 & (1.34) &  -0.04 &  (1.61) & 0.19 & (0.12) &  -0.51 & (0.18) &     2.60 & (7.04) \\[-0.0cm]
$\infty$ &   0.88 & (0.53) &   0.69 & (0.88) &   -0.14 & (0.81) &  -0.06 &  (0.99) & 0.21 & (0.12) &  -0.51 & (0.17) &    15.12 & (6.46) \\[0.1cm]
\hline
&&&&&&&&&&&&&&\\[-0.3cm]
& \multicolumn{14}{c}{\underline{Model$(\tau, \alpha)$}}   \\[-0.0cm]
&&&&&&&&&&&&&&\\[-0.3cm]
$\nu$ &  \multicolumn{2}{c}{$\hat\tau_0$} & \multicolumn{2}{c}{$\hat\tau_1$} & \multicolumn{2}{c}{$\hat\beta_0$} & \multicolumn{2}{c}{$\hat\beta_1$} & \multicolumn{2}{c}{$\hat\alpha_0$} &  \multicolumn{2}{c}{$\hat\alpha_1$} & \multicolumn{2}{c}{$\hat\nu_0$} \\[-0.0cm]
 \hline
&&&&&&&&&&&&&&\\[-0.3cm]
   0.00  &   0.80 & (0.09) &   0.60 & (0.13) &   0.00 &   ---  &   0.00 &    ---  &  0.21 & (0.06) &  -0.50 & (0.06) &     0.00 & (0.06) \\[-0.0cm]
   0.22  &   0.81 & (0.09) &   0.60 & (0.12) &   0.00 &   ---  &   0.00 &    ---  &  0.21 & (0.06) &  -0.50 & (0.06) &     0.22 & (0.09) \\[-0.0cm]
   0.41  &   0.80 & (0.08) &   0.59 & (0.10) &   0.00 &   ---  &   0.00 &    ---  &  0.20 & (0.06) &  -0.50 & (0.06) &     0.40 & (0.11) \\[-0.0cm]
   0.69  &   0.79 & (0.08) &   0.60 & (0.10) &   0.00 &   ---  &   0.00 &    ---  &  0.20 & (0.06) &  -0.50 & (0.06) &     0.70 & (0.17) \\[-0.0cm]
   1.10  &   0.80 & (0.09) &   0.60 & (0.09) &   0.00 &   ---  &   0.00 &    ---  &  0.20 & (0.06) &  -0.50 & (0.06) &     1.12 & (0.84) \\[-0.0cm]
   1.61  &   0.81 & (0.08) &   0.60 & (0.08) &   0.00 &   --- &   0.00 &    ---  &  0.20 & (0.07) &  -0.50 & (0.06) &     1.59 & (2.44) \\[-0.0cm]
$\infty$ &   0.82 & (0.05) &   0.62 & (0.06) &   0.00 &   ---  &   0.00 &    ---  &  0.22 & (0.05) &  -0.50 & (0.06) &    13.16 & (6.78) \\[0.1cm]
\hline
&&&&&&&&&&&&&&\\[-0.3cm]
& \multicolumn{14}{c}{\underline{Model$(\beta, \alpha)$}}   \\[-0.0cm]
&&&&&&&&&&&&&&\\[-0.3cm]
$\nu$ &  \multicolumn{2}{c}{$\hat\tau_0$} & \multicolumn{2}{c}{$\hat\tau_1$} & \multicolumn{2}{c}{$\hat\beta_0$} & \multicolumn{2}{c}{$\hat\beta_1$} &  \multicolumn{2}{c}{$\hat\alpha_0$} & \multicolumn{2}{c}{$\hat\alpha_1$} & \multicolumn{2}{c}{$\hat\nu_0$} \\[-0.0cm]
 \hline
&&&&&&&&&&&&&&\\[-0.3cm]
   0.00  &   0.00 &   ---  &   0.00 &   ---  &   0.88 & (0.11) &   0.03 &  (0.08) &  0.18 & (0.05) &  -0.52 & (0.05) &    -0.36 & (0.12) \\[-0.0cm]
   0.22  &   0.00 &   ---  &   0.00 &   ---  &   0.91 & (0.11) &   0.04 &  (0.08) &  0.18 & (0.06) &  -0.51 & (0.06) &    -0.06 & (0.15) \\[-0.0cm]
   0.41  &   0.00 &   ---  &   0.00 &   ---  &   0.93 & (0.13) &   0.05 &  (0.09) &  0.19 & (0.06) &  -0.50 & (0.06) &     0.21 & (0.21) \\[-0.0cm]
   0.69  &   0.00 &   ---  &   0.00 &   ---  &   0.98 & (0.13) &   0.06 &  (0.09) &  0.20 & (0.06) &  -0.50 & (0.06) &     0.72 & (0.35) \\[-0.0cm]
   1.10  &   0.00 &   ---  &   0.00 &   ---  &   1.03 & (0.14) &   0.07 &  (0.11) &  0.22 & (0.06) &  -0.50 & (0.07) &     1.83 & (4.33) \\[-0.0cm]
   1.61  &   0.00 &   ---  &   0.00 &   ---  &   1.18 & (0.10) &   0.08 &  (0.12) &  0.27 & (0.05) &  -0.50 & (0.07) &    15.16 & (6.52) \\[-0.0cm]
$\infty$ &   0.00 &   ---  &   0.00 &   ---  &   1.54 & (0.10) &   0.10 &  (0.14) &  0.37 & (0.05) &  -0.48 & (0.07) &    16.88 & (1.28) \\[0.1cm]
 \hline
&&&&&&&&&&&&&&\\[-0.2cm]
\multicolumn{15}{p{.95\textwidth}}{\footnotesize All numbers are rounded to two decimal places. For the models with fixed parameters, the ``estimated'' value shown is the value at which the parameter is fixed, and its standard error is then indicated by ``---''.}
\end{tabular}
\end{footnotesize}
\end{center}
\end{table}

Mirroring the case with no covariates (Section \ref{sec:nocov}), we find
that estimation is unstable when attempting to estimate $\tau$ and
$\beta$ coefficients simultaneously in $M(\tau,\beta,\alpha)$, {whereas
estimation is stable in both the true $M(\tau,\alpha)$ and the misspecified
 $M(\beta,\alpha)$} {models}. {In the latter,}  $\beta$
coefficients converge
consistently to values varying smoothly with $\nu$. 
Note that the
results are broadly similar for   smaller sample sizes of $n=500$ and
$n=100$ (see Appendix for details).

\begin{figure}[!htbp]
\begin{tabular}{c@{}c}
{\footnotesize\boldmath$\nu=0$} & {\footnotesize\boldmath$\nu=0.22$} \\
\includegraphics[width=0.48\textwidth, trim = 0.0cm 0.4cm 0.5cm 1.8cm, clip]{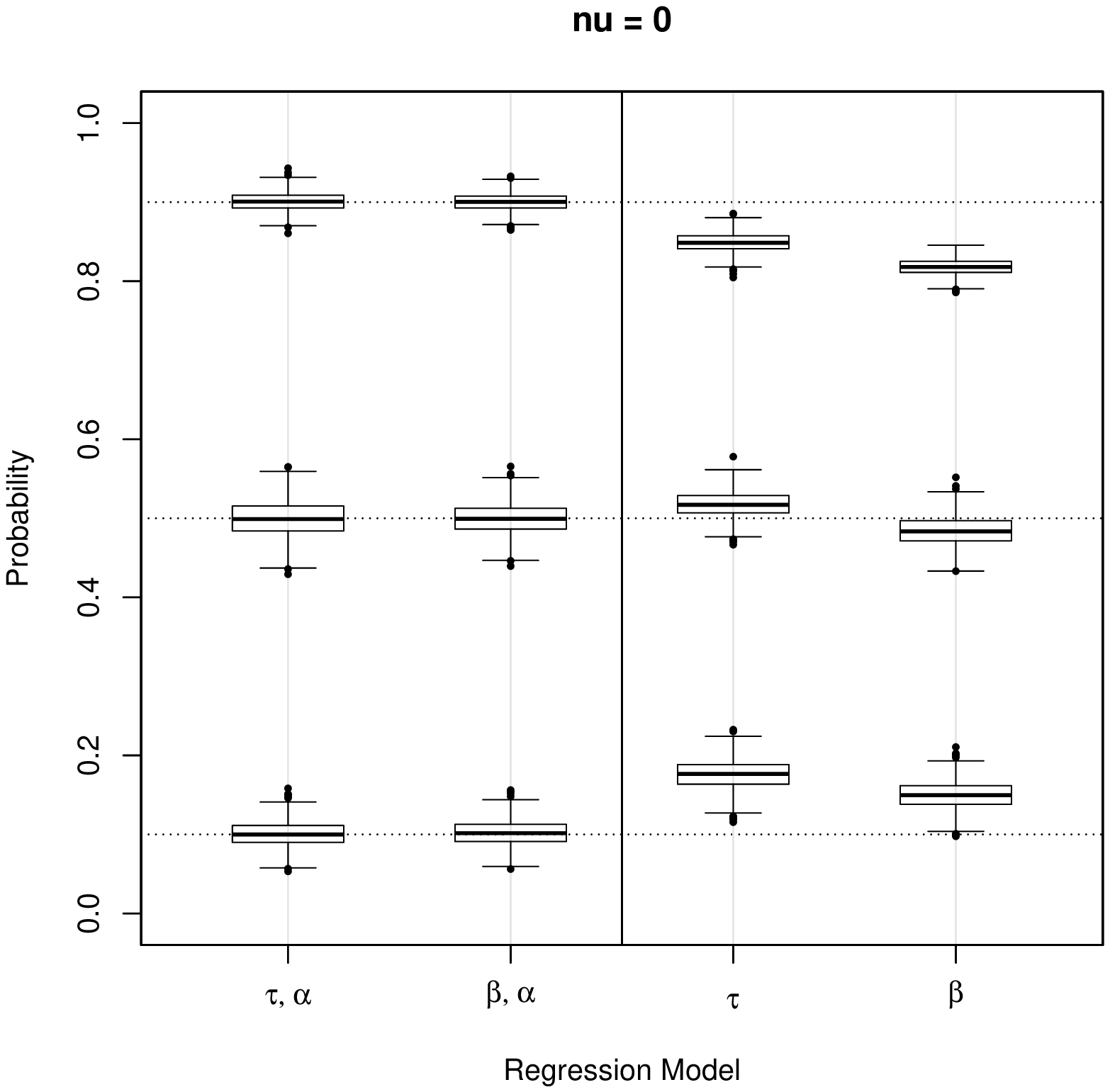} &
\includegraphics[width=0.48\textwidth, trim = 0.0cm 0.4cm 0.5cm 1.8cm, clip]{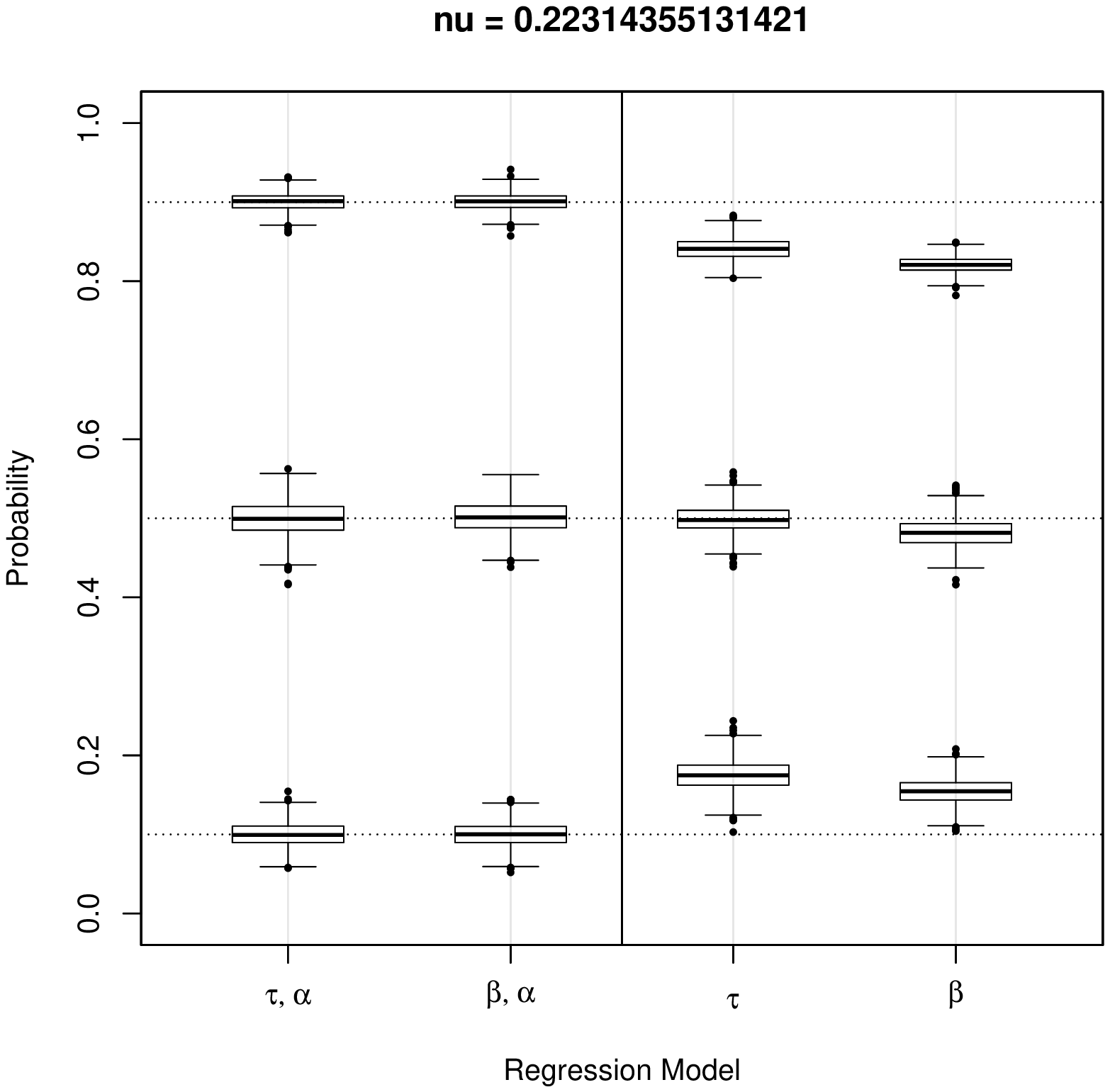}\\
{\footnotesize\boldmath$\nu=1.61$} & {\footnotesize\boldmath$\nu=\infty$} \\
\includegraphics[width=0.48\textwidth, trim = 0.0cm 0.4cm 0.5cm 1.8cm, clip]{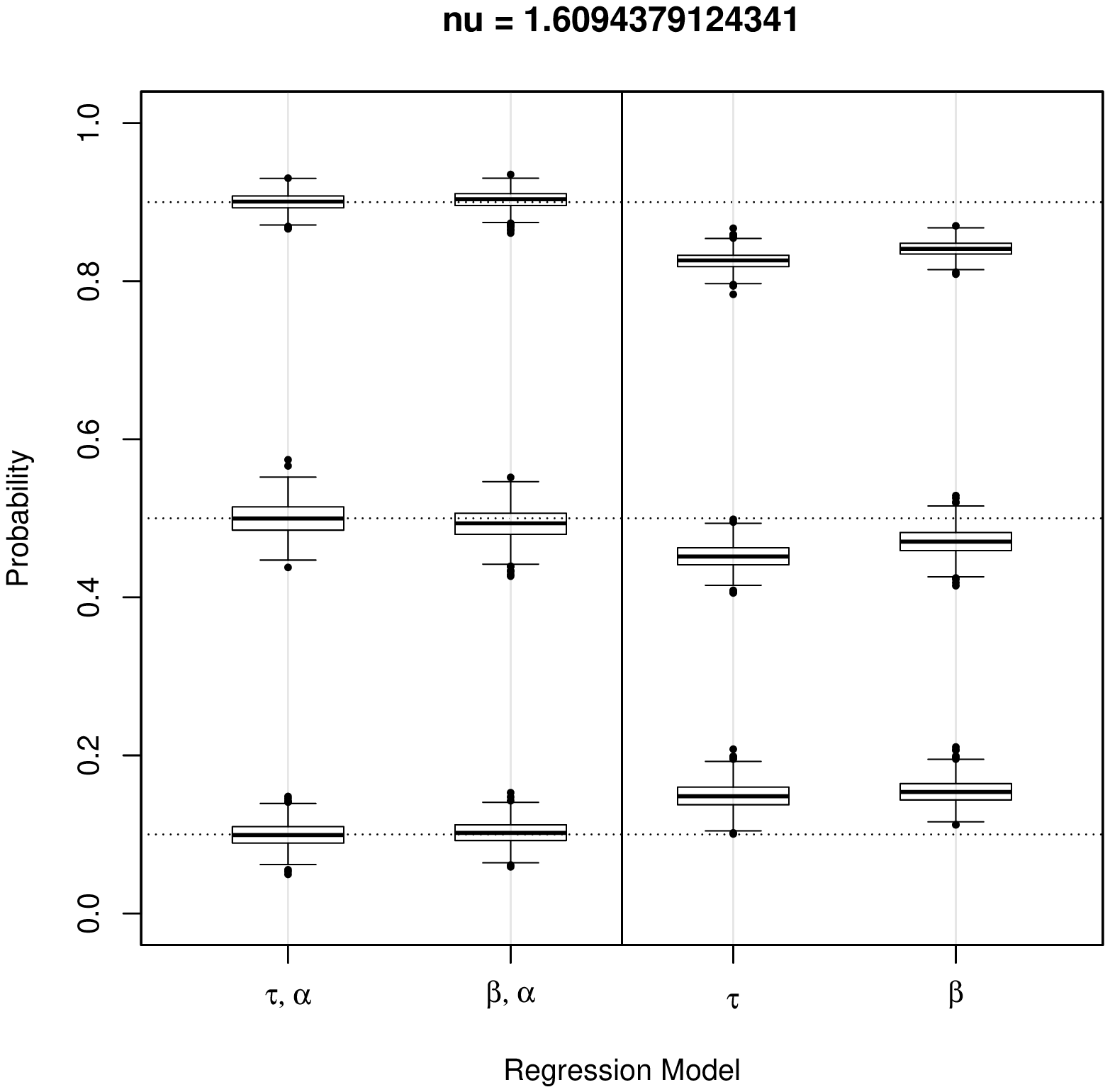} &
\includegraphics[width=0.48\textwidth, trim = 0.0cm 0.4cm 0.5cm 1.8cm, clip]{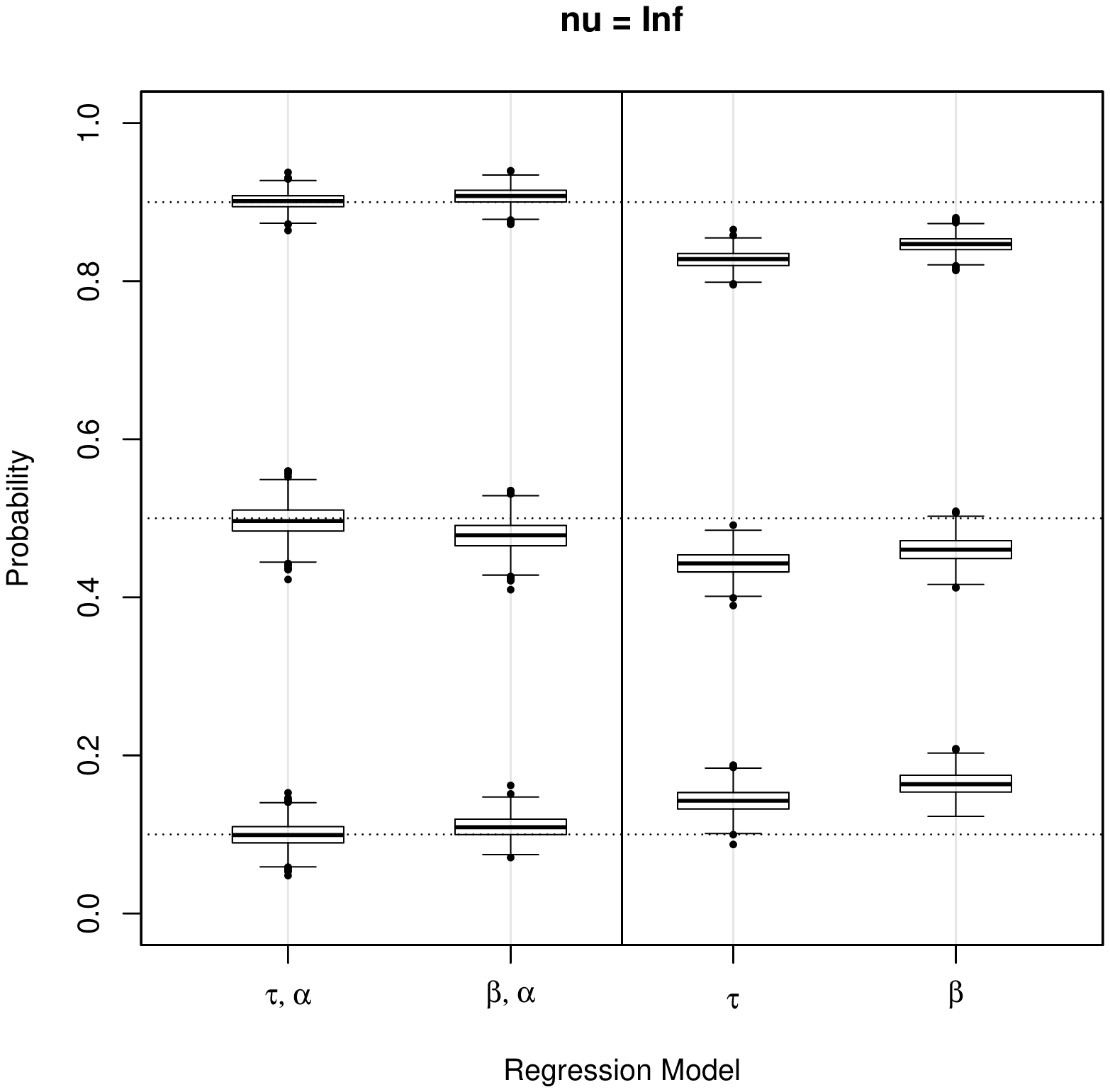}\\[0.7cm]
\end{tabular}
\caption{Boxplots of estimated baseline survivor probabilities
evaluated
at the true 90th, 50th, and 10th percentile times,  respectively (such
that the true probabilities are 0.1, 0.5, and 0.9),  vertically stacked
for each of four fitted models indicated by the $x$-axis labels.
\label{fig:prob}}
\end{figure}

{We now consider model fit} by inspecting the {estimated}
baseline survivor curves, i.e., the survivor curve for an individual with $X=0$ which we denote by $S_0(t)$. In particular, we focus on this estimated
baseline survivor function evaluated at three true quantiles, namely,
$Q_0(u)$, $u=0.1,0.5,0.9$,  since $\hat S_0(Q_0(u))$ is an estimate of the
probability $1-u$.
Boxplots of these estimates over simulation replicates arising from the
true model, $M(\tau,\alpha)$, and the misspecified model, $M(\beta,\alpha)$, are shown in Figure \ref{fig:prob}. 
We also display the estimates from two simpler (misspecified) models, $M(\tau)$ and $M(\beta)$, wherein $X$ has been dropped from the $\alpha$ component (specifically, $\alpha_1$ is set to zero in these simpler models).
{Clearly both $M(\tau,\alpha)$ and $M(\beta,\alpha)$ fit the data very well}
(apart from a little bias in $M(\beta,\alpha)$ when $\nu=\infty$), {i.e., the choice of using a $\tau$ or $\beta$ regression component does not alter the model fit much (again mirroring the findings of Section \ref{sec:nocov}). On the other hand,} when the $\alpha$
regression is dropped, the quality of the model fit decreases
considerably {as this represents a model misspecification in a much}
{stronger} {sense than switching from $M(\tau,\alpha)$ to
$M(\beta,\alpha)$.} 

%

\section{Data Analysis\label{sec:real}}

\subsection{Lung Cancer\label{sec:lung}}

We now consider our modelling framework in the context of a lung cancer
study which was the subject of a 1995 Queen's University Belfast PhD
thesis by P. Wilkinson (see \citet{burmac:2017}). This
study concerns 855 individuals who were diagnosed with lung cancer
between
1st October 1991 and 30th September 30 1992, who were then followed up
until 30th May 1993 (approximately 20\% of survival times were
right-censored). The main aim of this study was to investigate differences
between the following treatment groups: palliative care, surgery,
chemotherapy, radiotherapy, and a combined treatment of chemotherapy and
radiotherapy. In our analysis we take palliative care (which is a
non-curative  treatment providing pain relief) as the reference category.
Note that,  while various other covariates were captured for each
individual, our main aim here is to explore the flexibility of our general modelling scheme in the context of the treatment variable. 

{As discussed in Section \ref{sec:regression}, we advocate the use of $M(\beta,\alpha)$ and $M(\tau,\alpha)$ since they offer a flexible extension of the popular PH and AFT models (i.e., $M(\beta)$ and $M(\tau)$, respectively) in which the $\alpha$ coefficients indicate non-PH/non-AFT effects (see Section \ref{sec:interpretation}), and where the baseline distribution is selected via the parameter $\nu_0 = \log(\kappa+1)$. Thus, we fitted these two models, and their simpler PH and AFT counterparts, to the lung cancer data. We also fitted $M(\beta,\nu)$ and $M(\tau,\nu)$  for comparison, albeit we have argued in Section \ref{sec:interpretation} that these are perhaps somewhat less natural. These six fitted models are summarised in Table \ref{tab:lung}.}

We immediately see that the largest AIC/BIC values are associated with the
simpler single component (i.e., $\tau$ and $\beta$ only) models which
suggests that these models are not sufficiently flexible to capture
the  more
complex non-PH/non-AFT effects observed here. Although, in this
particular application, the AFT ($\tau$ only) model has lower AIC/BIC values than that of the PH ($\beta$ only) model, the fit can be greatly
improved  by modelling shape (either $\alpha$ or $\nu$) in addition
to scale. {Although the best-fitting model here is $M(\beta,\nu)$, the
difference is negligible compared with the models we favour,
$M(\beta,\alpha)$ and $M(\tau,\alpha)$. Interestingly, these latter two}
models have very close AIC/BIC values, indicating that the choice of
$\tau$ or $\beta$ component is not {at all} important here (in line
with the
findings of Section \ref{sec:simcov}). {The use of more than  two
regression components did not yield further improvements in fit  (models
not shown), and, moreover, estimation of such models tends to be  unstable
--- particularly}{, of course,} {those with two scale regression
components (see also
Section \ref{sec:sim}). Note that we have also avoided shape-only regression models, i.e., $M(\alpha)$, $M(\nu)$, and $M(\alpha,\nu)$, as, typically, models without scale components are not of interest, and, as we would expect, these models fit the current data very poorly indeed (with $\Delta_{\text{AIC}} > 600$).}

\begin{table}[!htbp]
\begin{center}
{
\caption{Summary of models fitted to lung cancer data\label{tab:lung}}
\smallskip
\begin{tabular}{c@{~~~~~}r@{~~~}r@{~~~}r@{~~~}r@{~~~}r@{~~~}r}
\hline
&&&&&&\\[-0.3cm]
Model                    & $M(\beta)$ & $M(\tau)$ & $M(\beta,\alpha)$ & $M(\tau,\alpha)$ & $M(\beta,\nu)$ & $M(\tau,\nu)$ \\[0.1cm]
\hline
&&&&&&\\[-0.3cm]
$\dim(\theta)$           &  7         &    7      & 11                &  11              &   11           &   11           \\[0.1cm]
$\ell(\hat\theta)$       & -1956.5    & -1943.5   & -1927.0           & -1926.6          & -1925.9        & -1930.3        \\[0.1cm]
$\Delta_{\text{AIC}}$    & 53.1       &  27.2     & 2.2               &  1.5             &   0.0          &   8.9          \\[0.1cm]
$\Delta_{\text{BIC}}$    &  34.1      &   8.2     & 2.2               &  1.5             &   0.0          &   8.9          \\[0.1cm]
\hline
\multicolumn{7}{p{.8\textwidth}}{\footnotesize $\ell(\hat\theta)$, the value of the log-likelihood; $\dim(\theta)$, the dimension of the model, i.e., number of parameters; $\Delta_{\text{AIC}}$, the AIC values for each model minus AIC$_{M(\beta,\nu)} = 3873.8$ (the lowest AIC in the set); $\Delta_{\text{BIC}}$, analogous to $\Delta_{\text{AIC}}$ where the lowest BIC is BIC$_{M(\beta,\nu)} = 3926.1$.}
\end{tabular}
}
\end{center}
\end{table}

%
%


\begin{figure}[!htb]
\begin{center}
\includegraphics[width=0.8\textwidth, trim = 0.0cm 0.4cm 0.5cm 1.8cm,
clip]{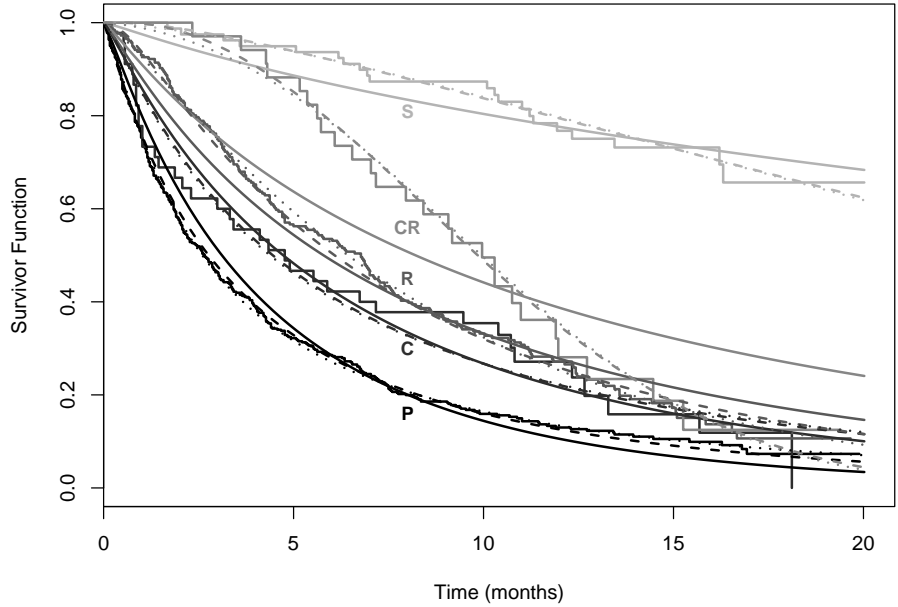}
\caption{Kaplan-Meier survivor curves (step, solid) for lung cancer treatment groups (P = palliative, C = chemotherapy, R = radiotherapy, CR =
chemotherapy and radiotherapy, and S = surgery)      with fitted curves
overlayed for $M(\beta)$ (solid), $M(\beta,\alpha)$  (dashed), and
$M(\beta,\nu)$ (dotted).\label{fig:kmfit}}
\end{center}
\end{figure}

{We now consider} the PH-APGW model, $M(\beta)$, and the two associated shape-regression extensions, $M(\beta,\alpha)$ and $M(\beta,\nu)$, {in more detail}. The advantage, in terms of
model fit, of shape regression components is clear from Figure
\ref{fig:kmfit}, while the $M(\beta,\alpha)$ and $M(\beta,\nu)$ models
themselves are virtually indistinguishable. Table \ref{tab:lung2} displays
the estimated regression coefficients. We can see that both $M(\beta)$ and
$M(\beta,\alpha)$ suggest a baseline distribution which is between a
log-logistic ($\nu=0$) and a Weibull ($\nu= 0.69$), while $M(\beta,\nu)$
assumes a separate baseline distribution for each treatment group.
Interestingly, in all three models, all shape parameters ($\nu$ and
$\alpha$) are positive which indicates that the hazards are increasing
with time in each treatment group (Table \ref{tab:pgwshapes}). While all
three models are in agreement when it comes to the overall effectiveness
of each treatment as viewed in terms of the scale coefficients (albeit
the chemotherapy effect is only statistically significant in
$M(\beta)$), the positive shape coefficients in  $M(\beta,\alpha)$ suggest
that the effectiveness of each treatment reduces to some extent over time
(see Section \ref{sec:interpretation}) -- especially in the case of the
combined treatment of chemotherapy and radiotherapy.

\begin{table}[!htb]
\begin{small}
\begin{center}
\caption{Selected lung cancer models\label{tab:lung2}}
\smallskip
\begin{tabular}{cr@{~~}cc@{~~~~}r@{~~}cr@{~~}cc@{~~~~}r@{~~}cr@{~~}c}
\hline
&&&&&&&&&&&&\\[-0.3cm]
& \multicolumn{2}{c}{Model$(\beta)$} && \multicolumn{4}{c}{Model$(\beta,\alpha)$} && \multicolumn{4}{c}{Model$(\beta,\nu)$} \\[0.1cm]
\cline{2-3}\cline{5-8}\cline{10-13}
&&&&&&&&&&&&\\[-0.3cm]
            & \multicolumn{2}{c}{Scale} && \multicolumn{2}{c}{Scale} & \multicolumn{2}{c}{Shape} && \multicolumn{2}{c}{Scale} & \multicolumn{2}{c}{Shape} \\
\hline
&&&&&&&&&&&&\\[-0.3cm]
Intercept      & -1.40 & (0.08) &&  -1.13 &  (0.09)  &  0.12 &  (0.07) &&  -1.04 &  (0.10)  &  0.21 &  (0.06)  \\[-0.0cm]
Palliative     &  0.00 &  ---   &&   0.00 &   ---    &  0.00 &   ---   &&   0.00 &   ---    &  0.00 &   ---    \\[-0.0cm]
Surgery        & -2.18 & (0.23) &&  -4.77 &  (0.97)  &  1.06 &  (0.28) &&  -3.96 &  (0.66)  &  0.55 &  (0.15)  \\[-0.0cm]
Chemo          & -0.38 & (0.17) &&  -0.55 &  (0.33)  &  0.13 &  (0.18) &&  -0.60 &  (0.36)  &  0.11 &  (0.13)  \\[-0.0cm]
Radio          & -0.56 & (0.09) &&  -1.46 &  (0.21)  &  0.52 &  (0.11) &&  -1.48 &  (0.19)  &  0.36 &  (0.06)  \\[-0.0cm]
C$+$R          & -0.86 & (0.20) &&  -5.13 &  (0.96)  &  1.50 &  (0.22) &&  -3.57 &  (0.60)  &  0.82 &  (0.13)  \\[0.2cm]
$\hat\alpha_0$ &  0.15 & (0.08) &&   \multicolumn{4}{c}{$\ast$}   &&   \multicolumn{4}{c}{0.27~~(0.07)}   \\[-0.0cm]
$\hat\nu_0$    &  0.46 & (0.06) &&   \multicolumn{4}{c}{0.35~~(0.05)}  &&   \multicolumn{4}{c}{$\ast$}    \\[0.1cm]
\hline
&&&&&&&&&&&&\\[-0.2cm]
\multicolumn{13}{p{.95\textwidth}}{\footnotesize The $\ast$ symbol indicates that the shape parameter
already appears as the intercept in the shape regression component.}\\[-0.3cm]
\end{tabular}
\end{center}
\end{small}
\end{table}

\begin{figure}[!tbp]
\begin{center}
\includegraphics[width=0.9\textwidth, trim = 0.0cm 0.4cm 0.5cm 0.8cm, clip]{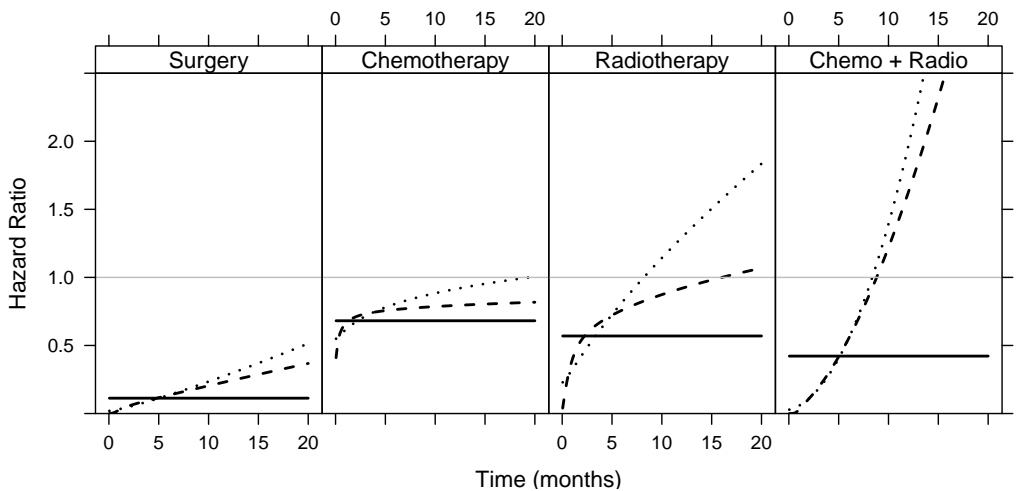}
\caption{Hazard ratios for each lung cancer treatment relative to palliative care for
$M(\beta)$ (solid), $M(\beta,\alpha)$ (dashed), and $M(\beta,\nu)$
(dotted). Line of equality (grey) also shown.\label{fig:hr}}
\end{center}
\end{figure}

The hazard
 ratios for the models {in} {Table \ref{tab:lung2}} are shown in
Figure \ref{fig:hr} where those of
 $M(\beta,\alpha)$ and $M(\beta,\nu)$ are {quite} similar. They
suggest that while the various treatments reduce the hazard  in the first
few months, their effect is weakened over time and, perhaps, even  become
inferior to palliative care in the longer term  (however, note that very
few individuals remain in the sample beyond 15 months).  Clearly SPR
models, such as $M(\beta)$,  cannot account for covariate effects of
this sort.

It is worth
highlighting the fact that the basic
findings
 here are qualitatively similar to those of \citet{burmac:2017} who
 analysed this lung cancer dataset using $M_{\kappa=1}(\beta,\alpha)$,
 i.e., a Weibull MPR model. However, the framework of the current paper
 permits us to consider a much wider range of model structures and
 distributions in which $M_{\kappa=1}(\beta,\alpha)$ appears as a special
 case. In particular, $M(\beta,\alpha)$ from Table \ref{tab:lung2} yields
 a 95\% confidence interval for $\kappa$, $[0.28, 0.57]$, which does not
 support the Weibull ($\kappa=1$) baseline distribution. Furthermore,
 $\text{AIC}_{M_{\kappa=1}(\beta,\alpha)}-\text{AIC}_{M(\beta,\alpha)} =
 57.6$, and we can confirm that the improvement in quality of fit is most
 evident in the palliative care group (which
 $M_{\kappa=1}(\beta,\alpha)$ does not capture so well).  Thus,
although
 the basic findings are unaltered in this particular application, the APGW
 MPR approach yields a better solution in which uncertainty in selecting
 the baseline distribution is accounted for. Of course, the APGW MPR model
 will readily adapt to other applications which might differ significantly
 (both qualitatively and quantitatively) from
 $M_{\kappa=1}(\beta,\alpha)$.

%


\subsection{Melanoma\label{sec:melanoma}}

The Eastern Cooperative Oncology Group (ECOG) trial ``EST 1684'' was a
randomized controlled trial to investigate the adjuvant (i.e.,
post-surgery) chemotherapy drug ``IFN$\alpha$-2b'' in treating melanoma
\citep{kirkwood:1996}. The outcome variable was relapse-free survival,
i.e., time from randomization until the earlier of cancer relapse or
death. Patients were recruited to the study between 1984 and 1990, and the
study ended in 1993. In total, 284 patients were recruited of which 140
were assigned to the control group, and 144 were assigned to the treatment
group. This dataset is available in the \texttt{R} package
\texttt{smcure} \citep{chaoetal:2012}, and variations of it have appeared in the cure model literature \citep{chen:1999,ibrahirmetal:2001}.

\begin{figure}[!htb]
\begin{center}
\includegraphics[width=0.8\textwidth, trim = 0.0cm 0.4cm 0.5cm 1.8cm,
clip]{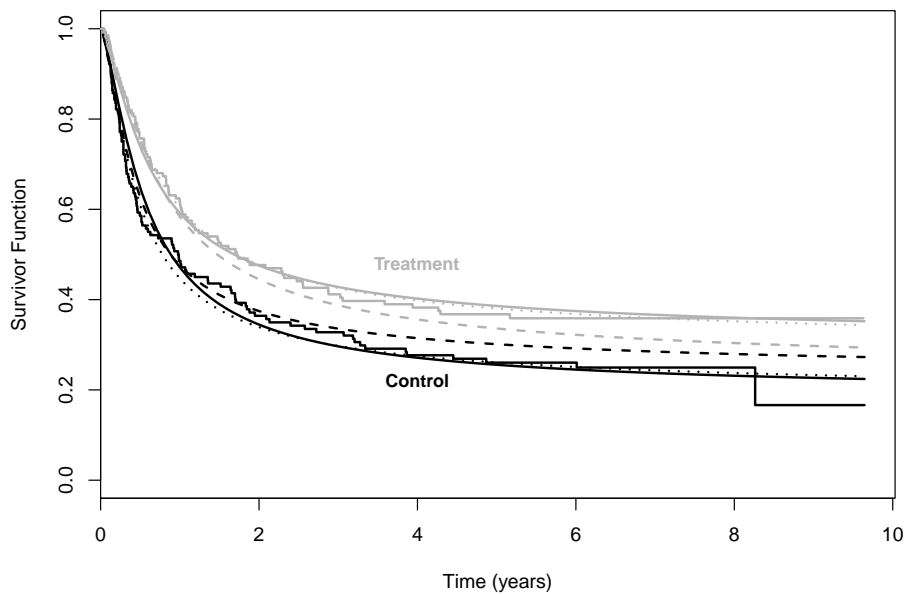}
\caption{Kaplan-Meier survivor curves (step, solid) for melanoma control and treatment groups with fitted curves
overlayed for $M(\beta)$ (solid), $M(\tau)$  (dashed), and
$M(\tau,\nu)$ (dotted).\label{fig:kmfitmel}}
\end{center}
\end{figure}

As in Section \ref{sec:lung}, we fitted the following models: $M(\beta)$,
$M(\tau)$, $M(\beta,\alpha)$, $M(\tau,\alpha)$, $M(\beta,\nu)$, and
$M(\tau,\nu)$; the results are summarised in Table \ref{tab:melanoma}. In
this case, the two-component models do not provide a large improvement
{over}
the one-component models, and $M(\tau)$ has the lowest AIC and the
second-lowest BIC ($M(\tau,\nu)$ has the lowest BIC); the AIC and BIC
values for $M(\beta)$ are not much larger than {for} $M(\tau)$. The
models
$M(\beta)$, $M(\tau)$, and $M(\tau,\nu)$ are compared to the Kaplan-Meier
(KM) curves in Figure \ref{fig:kmfitmel}. The fitted $M(\beta)$ and
$M(\tau,\nu)$ curves are similar, and are close to the KM curves. The
fitted $M(\tau)$ curves converge later in time, which, visually, look
worse compared to the KM curves. However, note that there is very little
data in the right tail so that converging curves are plausible when viewed
with the level of uncertainty in the tail.

\begin{table}[!htbp]
\begin{center}
\caption{Summary of models fitted to melanoma data\label{tab:melanoma}}
\smallskip
\begin{tabular}{c@{~~~~~}r@{~~~}r@{~~~}r@{~~~}r@{~~~}r@{~~~}r}
\hline
&&&&&&\\[-0.3cm]
Model                    & $M(\beta)$ & $M(\tau)$ & $M(\beta,\alpha)$ & $M(\tau,\alpha)$ & $M(\beta,\nu)$ & $M(\tau,\nu)$ \\[0.1cm]
\hline
&&&&&&\\[-0.3cm]
$\dim(\theta)$           & 4          & 4         & 5                 & 5                & 5              & 5      \\[0.1cm]
$\ell(\hat\theta)$       & -368.8     & -368.0    & -367.7            & -367.9           & -368.2         & -366.6 \\[0.1cm]
$\Delta_{\text{AIC}}$    & 2.4        & 0.8       & 2.1               & 2.6              & 3.2            & 0.0    \\[0.1cm]
$\Delta_{\text{BIC}}$    & 1.6        & 0.0       & 5.0               & 5.5              & 6.1            & 2.9    \\[0.1cm]
\hline
\multicolumn{7}{p{.8\textwidth}}{\footnotesize $\ell(\hat\theta)$, the value of the log-likelihood; $\dim(\theta)$, the dimension of the model, i.e., number of parameters; $\Delta_{\text{AIC}}$, the AIC values for each model minus AIC$_{M(\tau,\nu)} = 743.2$ (the lowest AIC in the set); $\Delta_{\text{BIC}}$, analogous to $\Delta_{\text{AIC}}$ where the lowest BIC is BIC$_{M(\tau)} = 758.6$.}
\end{tabular}
\end{center}
\end{table}

The {parameter estimates} for $M(\beta)$, $M(\tau)$, and
$M(\tau,\nu)$ are shown in
Table \ref{tab:melanoma2}. Firstly note that the scale coefficient of
treatment   is negative which indicates that treatment improves survival.
Furthermore,  for $M(\beta)$ and $M(\tau)$, note that the parameter
$\kappa = \exp(\nu_0)-1$ {is} negative, and, similarly, for
$M(\tau,\nu)$,
the {estimated} $\kappa$ is negative in each treatment group. Thus,
all three models
point towards a cure proportion.  The fact that the power shape parameter,
$\gamma = \exp(\alpha)$, is greater than one for all three models means
that the hazard function has an up-then-down shape. This type of hazard
function is commonly observed in {the} cure literature since
the population
becomes increasingly composed of cured individuals (i.e., zero hazard) over time.

\begin{table}[!htb]
\begin{small}
\begin{center}
\caption{Selected melanoma models\label{tab:melanoma2}}
\smallskip
\begin{tabular}{cr@{~~}cc@{~~~~}r@{~~}cr@{~~}r@{~~}cr@{~~}c}
\hline
&&&&&&&&&&\\[-0.3cm]
& \multicolumn{2}{c}{Model$(\beta)$} && \multicolumn{2}{c}{Model$(\tau)$} && \multicolumn{4}{c}{Model$(\tau,\nu)$} \\[0.1cm]
\cline{2-3}\cline{5-6}\cline{8-11}
&&&&&&&&&&\\[-0.3cm]
            & \multicolumn{2}{c}{Scale} && \multicolumn{2}{c}{Scale} && \multicolumn{2}{c}{Scale} & \multicolumn{2}{c}{Shape} \\
\hline
&&&&&&&&&&\\[-0.3cm]
Intercept      &   0.60  & (0.16) &&   0.65  &  (0.14)  &&   0.67  &  (0.15)  &  -0.45  &  (0.05)  \\[-0.0cm]
Control        &   0.00  &  ---   &&   0.00  &   ---    &&   0.00  &   ---    &   0.00  &   ---    \\[-0.0cm]
Treatment      &  -0.36  & (0.14) &&  -0.52  &  (0.20)  &&  -0.53  &  (0.19)  &  -0.14  &  (0.08)  \\[0.2cm]
$\hat\alpha_0$ &   0.36  & (0.07) &&   0.44  &  (0.08)  &&   \multicolumn{4}{c}{0.45~~(0.08)}   \\[-0.0cm]
$\hat\nu_0$    &  -0.73  & (0.11) &&  -0.51  &  (0.05)  &&   \multicolumn{4}{c}{$\ast$}    \\[0.1cm]
\hline
&&&&&&&&&&\\[-0.2cm]
\multicolumn{11}{p{.8\textwidth}}{\footnotesize The $\ast$ symbol indicates that the shape parameter
already appears as the intercept in the shape regression component.}\\[-0.3cm]
\end{tabular}
\end{center}
\end{small}
\end{table}

\begin{table}[!htb]
\begin{small}
\begin{center}
\caption{Estimated melanoma cure proportions with 95\% confidence intervals\label{tab:melanoma3}}
\smallskip
\begin{tabular}{c@{\qquad\quad}c@{\qquad}c@{\qquad}c}
\hline
&&&\\[-0.3cm]
Model           &   Treatment         & Control & Difference \\
\hline
&&&\\[-0.3cm]
$M(\beta)$      &   0.30~(0.21,0.39)  & 0.18~(0.10,0.26) & 0.12~(0.03,0.21) \\[0.1cm]
$M(\tau)$       &   0.22~(0.15,0.30)  & $=$ Treatment    & --- \\[0.1cm]
$M(\tau,\nu)$   &   0.24~(0.18,0.39)  & 0.11~(0.09,0.27) & 0.12~(-0.02,0.23) \\[0.1cm]
\hline
&&&\\[-0.3cm]
\multicolumn{4}{p{.7\textwidth}}{\footnotesize $p_{\text{Difference}} = p_{\text{Treatment}} - p_{\text{control}}$.}\\[-0.3cm]
\end{tabular}
\end{center}
\end{small}
\end{table}

When {$-1 < \kappa < 0$}, the APGW-MPR cure proportion is given
by\linebreak
$\exp\{\lambda (\kappa+1)/\kappa\} > 0$, i.e., the cure proportion depends
on $\kappa$ and $\lambda$. Therefore, $M(\beta)$ and $M(\tau,\nu)$ suggest
that the cure proportion depends on treatment, while $M(\tau)$ suggests
that it does not. The estimated cure proportions, along with 95\%
confidence intervals, are shown in Table \ref{tab:melanoma3}. Had we fixed
to a log-logistic baseline (i.e., $\kappa=0$), the resulting
$M_{\kappa=0}(\beta)$ and $M_{\kappa=0}(\tau)$ models (not shown) provide
an extremely poor fit to the data. This is noteworthy as even the
heaviest-tail non-cure APGW model is not supported by the data (and, of
course, a Weibull baseline is worse still). The heaviness of tail here can
only be supported {within the APGW family by a cure model}.

\section{Discussion\label{sec:disc}}

Our proposed APGW-MPR modelling framework {is} highly flexible and
{can} adapt readily to a wide variety of applications in survival
analysis and reliability. In particular,
this framework includes the practically important AFT and PH
models, and
generalises them through shape regression components. Furthermore, the
 APGW baseline model covers the primary shapes of hazard
function (constant, increasing, decreasing, up-then-down,
down-then-up) within some
of the most popular survival  distributions (log-logistic, Burr type
XII, Weibull,
Gompertz) using only two shape parameters. 

In practice, the full four-component APGW-MPR model is likely to be more
flexible than is required for most purposes. {In fact}, {we}
suggest that covariates should appear via just one
scale-type
component
($\tau$ or $\beta$), along with the $\alpha$ shape component which permits
survivor functions with differing shapes and indicates departures from
more basic AFT or PH effects, while {$\nu = \log (\kappa+1)$} is a
covariate-independent
parameter which allows us to choose among  distributions within one
unified framework. {We} have found that the scale-type
parameters ($\tau$ and $\beta$) are highly intertwined in the sense that
they cannot be estimated simultaneously within the same model reliably,
and are highly correlated.
{This} is
true
across the full range of distributions (varying $\nu$), going well
beyond
the well-known Weibull case in which the two {scale components are equivalent}. The implication of this is that, in terms of performance gain,
the movement from AFT to PH modelling (or vice versa)
might not be
very large, whereas we
have found that modelling the shape is a more fruitful alteration to
the
regression specification. 

 Finally, the perspective of this paper has been to investigate survival
 modelling generally, to cover some of the most popular models, and
to discover
 some of the better modelling choices that can be made within this
 framework. Although we have developed these ideas in a fully
 parametric context, non-parametric equivalents, while possible, are
 beyond the scope of the present paper (but are investigated in a separate
 line of work \citep{burerikpip:2018}). However, it is worth highlighting
 that perhaps too much emphasis is placed on non-/semi-parametric
 approaches in survival analysis whereby undue weight is attached to the
 flexibility of the baseline distribution in comparison to the
 flexibility of its regression structure. 
 {Our} general approach to survival
 modelling provides a framework within which one can consider the most
 important components of survival modelling (including which might
potentially be modelled non-parametrically), and we believe that this
insight can lead to better modelling practice in general.

\section*{Acknowledgements}
The first author would like to thank the Irish Research Council (www.research.ie) for supporting this work (New Foundations award).

\bibliographystyle{apalike}
\bibliography{refs}

\begin{thebibliography}{}

\bibitem[Bagdonavi\c{c}ius and Nikulin, 2002]{bagnik:2002}
Bagdonavi\c{c}ius, V. and Nikulin, M. (2002).
\newblock {\em Accelerated Life Model; Modeling and Statistical Analysis}.
\newblock Chapman \& Hall/CRC, Boca Raton, FL.

\bibitem[Box and Cox, 1964]{boxcox:1964}
Box, G. and Cox, D. (1964).
\newblock An analysis of transformations (with discussion).
\newblock {\em Journal of the Royal Statistical Society Series B}, 26:211--252.

\bibitem[Burke et~al., 2018]{burerikpip:2018}
Burke, K., Eriksson, F., and Pipper, C.~B. (2018).
\newblock Semiparametric multi-parameter regression survival modelling.
\newblock {\em Submitted}.

\bibitem[Burke and MacKenzie, 2017]{burmac:2017}
Burke, K. and MacKenzie, G. (2017).
\newblock Multi-parameter regression survival modelling: An alternative to
  proportional hazards.
\newblock {\em Biometrics}, 73:678--686.

\bibitem[Chao et~al., 2012]{chaoetal:2012}
Chao, C., Yubo, Z., Yingwei, P., and Jiajia, Z. (2012).
\newblock {\em smcure: fit semiparametric mixture cure models}.
\newblock R package version 2.0.

\bibitem[Chen et~al., 1999]{chen:1999}
Chen, M., Ibrahim, J., and Sinha, D. (1999).
\newblock A new {B}ayesian model for survival data with a surviving fraction.
\newblock {\em J. Am. Statist. Ass.}, 94:909--919.

\bibitem[Chen and Jewell, 2001]{chenjewell:2001}
Chen, Y. and Jewell, N. (2001).
\newblock On a general class of semiparametric hazards regression models.
\newblock {\em Biometrika}, 88:687--702.

\bibitem[Chen, 2000]{chen:2000}
Chen, Z. (2000).
\newblock A new two-parameter lifetime distribution with bathtub shape or
  increasing failure rate function.
\newblock {\em Statistical Probability Letters}, 49:155--161.

\bibitem[Cox and Matheson, 2014]{coxmath:2014}
Cox, C. and Matheson, M. (2014).
\newblock A comparison of the generalized gamma and exponentiated {W}eibull
  distributions.
\newblock {\em Statistics in Medicine}, 33:3772--3780.

\bibitem[Dimitrakopoulou et~al., 2007]{dimetal:2007}
Dimitrakopoulou, T., Adamidis, K., and Loukas, S. (2007).
\newblock A lifetime distribution with an upside-down bathtub-shaped hazard
  function.
\newblock {\em IEEE Transactions on Reliability}, 56:308--311.

\bibitem[Ibrahim et~al., 2001]{ibrahirmetal:2001}
Ibrahim, J., Chen, M., and Sinha, D. (2001).
\newblock Bayesian semiparametric models for survival data with a cure
  fraction.
\newblock {\em Biometrics}, 57:383--388.

\bibitem[Jones and Noufaily, 2015]{jonesnouf:2015}
Jones, M. and Noufaily, A. (2015).
\newblock Log-location-scale-log-concave distributions for survival and
  reliability analysis.
\newblock {\em Electronic Journal of Statistics}, 9:2732--2750.

\bibitem[Jones et~al., 2018]{jnb:2018}
Jones, M., Noufaily, A., and Burke, K. (2018).
\newblock A bivariate power generalized {W}eibull distribution: a flexible
  parametric model for survival analysis.
\newblock {\em Submitted}.

\bibitem[Kalbfleisch and Prentice, 2002]{kalbprent:2002}
Kalbfleisch, J. and Prentice, R. (2002).
\newblock {\em The Statistical Analysis of Failure Time Data}.
\newblock Wiley, Hoboken, NJ., 2 edition.

\bibitem[Kirkwood et~al., 1996]{kirkwood:1996}
Kirkwood, J., Strawderman, M., Ernstoff, M., Smith, T., Borden, E., and Blum,
  R. (1996).
\newblock Interferon alfa-2b adjuvant therapy of high-risk resected cutaneous
  melanoma: the {E}astern {C}ooperative {O}ncology {G}roup {T}rial {EST} 1684.
\newblock {\em J. Clinical Oncology}, 14:7--17.

\bibitem[Matheson et~al., 2017]{mathetal:2017}
Matheson, M., Mu\~{n}oz, A., and Cox, C. (2017).
\newblock Describing the flexibility of the generalized gamma and related
  distributions.
\newblock {\em Journal of Statistical Distributions and Applications},
  4:Article~15.

\bibitem[Nadarajah and Haghighi, 2011]{nadarhagh:2011}
Nadarajah, S. and Haghighi, F. (2011).
\newblock An extension of the exponential distribution.
\newblock {\em Statistics}, 45:543--558.

\bibitem[Nikulin and Haghighi, 2009]{nikhagh:2009}
Nikulin, M. and Haghighi, F. (2009).
\newblock On the power generalized {W}eibull family: model for cancer censored
  data.
\newblock {\em Metron}, 67:75--86.

\bibitem[Tsodikov et~al., 2003]{tsodikovetal:2003}
Tsodikov, A., Ibrahim, J., and Yakovlev, A. (2003).
\newblock Estimating cure rates from survival data: an alternative to
  two-component mixture models.
\newblock {\em Journal of the American Statistical Association}, 98:1063--1078.

\bibitem[Xie et~al., 2002]{xieetal:2002}
Xie, M., Tang, Y., and Goh, T. (2002).
\newblock A modified {W}eibull extension with bathtub-shaped failure rate
  function.
\newblock {\em Reliability Engineering and System Safety}, 73:279--285.

\bibitem[Yeo and Johnson, 2000]{yeojohnson:2000}
Yeo, I. and Johnson, R. (2000).
\newblock A new family of power transformations to improve normality or
  symmetry.
\newblock {\em Biometrika}, 87:954--959.

\bibitem[Zeng and Lin, 2007]{zenglin:2007}
Zeng, Z. and Lin, D. (2007).
\newblock Maximum likelihood estimation in semiparametric regression models
  with censored data (with discussion).
\newblock {\em Journal of the Royal Statistical Society Series B}, 69:507--564.

\end{thebibliography}

\newpage
\centerline{\sc{Appendix}}
\medskip

\appendix
\section{The (A)PGW Distribution as a Transformation Model}

Linear transformation models  (c.f.\ \citet[sec.~7.5]{kalbprent:2002};
\citet{zenglin:2007})
are concerned with c.h.f.'s of the form
\begin{equation}
H_T(t) = w(\theta H_0(t)) \label{transformation-model} \tag{A:1}
\end{equation}
where $\theta>0$ depends log-linearly on covariates and both the
transformation function  $w$ and baseline function
$H_0$ are c.h.f.'s.
It is easy to see that the (A)PGW c.h.f.'s $H_N$ and $H_A$ are of
the form (\ref{transformation-model}). Either is  a transformation model
with
$H_0$ the Weibull c.h.f.\ $t^\gamma$; in terms of the overall model,
 whenever the Weibull
is used as baseline,   $\theta^{-1/\gamma}$ is
the
horizontal scale parameter so that when covariates are included only in
it,  the transformation model is
an accelerated failure time model.  The transformation $w$ is a version of
the Box-Cox
transformation given, in the simpler case of $H_N$,  by $w(y) =
(1+y)^\kappa-1$ \citep{boxcox:1964,yeojohnson:2000}.

If $Y$ is the lifetime random variable following the
transformation model with, for simplicity, c.h.f.\ $H_N$, then the
model can also be written in the form
$\gamma \log Y = -\log\theta +\log E$
where
$E$ follows the distribution with c.h.f.\  $(1+y)^\kappa-1$.


\section{Simulation Studies with Smaller Samples}

This section displays analogous  results to those of Table 4 from the main paper
but for $n=500$ and $n=100$ respectively.

\begin{table}[htbp]
\begin{center}
\caption{Median and standard {error} (in brackets) of estimates when $n=500$}
\begin{footnotesize}
\begin{tabular}{c@{\qquad}r@{~}c@{~~}r@{~}c@{~~}r@{~}c@{~~}r@{~}c@{~~}r@{~}c@{~~}r@{~}c@{~~}r@{~}c}
\hline
&&&&&&&&&&&&&&\\[-0.3cm]
& \multicolumn{14}{c}{\underline{Model$(\tau, \beta, \alpha)$}} \\[-0.0cm]
&&&&&&&&&&&&&&\\[-0.3cm]
$\nu$ &  \multicolumn{2}{c}{$\hat\tau_0$} & \multicolumn{2}{c}{$\hat\tau_1$} & \multicolumn{2}{c}{$\hat\beta_0$} & \multicolumn{2}{c}{$\hat\beta_1$} & \multicolumn{2}{c}{$\hat\alpha_0$} &  \multicolumn{2}{c}{$\hat\alpha_1$} & \multicolumn{2}{c}{$\hat\nu_0$} \\[-0.0cm]
 \hline
&&&&&&&&&&&&&&\\[-0.3cm]
   0.00  &   1.06 & (1.57) &   0.57 & (1.63) &   -0.25 & (1.73) &    0.01 & (1.13) &   0.23 & (0.21) &   -0.49 & (0.32) &    0.07 & (1.02) \\[-0.0cm]
   0.22  &   1.30 & (1.62) &   0.45 & (2.66) &   -0.52 & (1.95) &    0.21 & (2.13) &   0.27 & (0.24) &   -0.53 & (0.37) &    0.27 & (0.73) \\[-0.0cm]
   0.41  &   1.05 & (1.73) &   0.81 & (3.17) &   -0.19 & (2.11) &   -0.03 & (2.72) &   0.27 & (0.24) &   -0.50 & (0.34) &    0.35 & (0.78) \\[-0.0cm]
   0.69  &   0.88 & (2.12) &   0.45 & (4.19) &   -0.15 & (2.60) &    0.19 & (3.70) &   0.21 & (0.26) &   -0.51 & (0.28) &    0.65 & (4.26) \\[-0.0cm]
   1.10  &   0.43 & (1.51) &   0.26 & (2.93) &    0.33 & (1.91) &    0.04 & (2.72) &   0.15 & (0.16) &   -0.49 & (0.23) &    1.53 & (7.11) \\[-0.0cm]
   1.61  &   0.53 & (1.09) &   0.40 & (1.98) &    0.25 & (1.47) &   -0.01 & (2.03) &   0.19 & (0.16) &   -0.50 & (0.24) &   14.75 & (7.06) \\[-0.0cm]
$\infty$ &   0.91 & (0.72) &   0.68 & (1.26) &   -0.15 & (1.07) &   -0.02 & (1.41) &   0.19 & (0.16) &   -0.51 & (0.24) &   15.43 & (6.48) \\[0.1cm]
\hline
&&&&&&&&&&&&&&\\[-0.3cm]
& \multicolumn{14}{c}{\underline{Model$(\tau, \alpha)$}}   \\[-0.0cm]
&&&&&&&&&&&&&&\\[-0.3cm]
$\nu$ &  \multicolumn{2}{c}{$\hat\tau_0$} & \multicolumn{2}{c}{$\hat\tau_1$} & \multicolumn{2}{c}{$\hat\beta_0$} & \multicolumn{2}{c}{$\hat\beta_1$} & \multicolumn{2}{c}{$\hat\alpha_0$} &  \multicolumn{2}{c}{$\hat\alpha_1$} & \multicolumn{2}{c}{$\hat\nu_0$} \\[-0.0cm]
 \hline
&&&&&&&&&&&&&&\\[-0.3cm]
   0.00  &   0.79 & (0.13) &   0.60 & (0.19) &   0.00 &   ---  &   0.00 &   ---  &   0.20 & (0.09) &   -0.50 & (0.09) &    0.00 & (0.09) \\[-0.0cm]
   0.22  &   0.80 & (0.12) &   0.59 & (0.16) &   0.00 &   ---  &   0.00 &   ---  &   0.20 & (0.09) &   -0.50 & (0.09) &    0.22 & (0.13) \\[-0.0cm]
   0.41  &   0.79 & (0.12) &   0.58 & (0.15) &   0.00 &   ---  &   0.00 &   ---  &   0.20 & (0.09) &   -0.50 & (0.08) &    0.42 & (0.18) \\[-0.0cm]
   0.69  &   0.80 & (0.13) &   0.58 & (0.14) &   0.00 &   ---  &   0.00 &   ---  &   0.20 & (0.09) &   -0.50 & (0.08) &    0.72 & (0.30) \\[-0.0cm]
   1.10  &   0.79 & (0.13) &   0.60 & (0.12) &   0.00 &   ---  &   0.00 &   ---  &   0.20 & (0.09) &   -0.51 & (0.09) &    1.13 & (1.73) \\[-0.0cm]
   1.61  &   0.80 & (0.11) &   0.60 & (0.11) &   0.00 &   ---  &   0.00 &   ---  &   0.20 & (0.09) &   -0.50 & (0.09) &    1.62 & (4.26) \\[-0.0cm]
$\infty$ &   0.84 & (0.07) &   0.62 & (0.08) &   0.00 &   ---  &   0.00 &   ---  &   0.24 & (0.08) &   -0.51 & (0.09) &   13.84 & (6.97) \\[0.1cm]
\hline
&&&&&&&&&&&&&&\\[-0.3cm]
& \multicolumn{14}{c}{\underline{Model$(\beta, \alpha)$}}   \\[-0.0cm]
&&&&&&&&&&&&&&\\[-0.3cm]
$\nu$ &  \multicolumn{2}{c}{$\hat\tau_0$} & \multicolumn{2}{c}{$\hat\tau_1$} & \multicolumn{2}{c}{$\hat\beta_0$} & \multicolumn{2}{c}{$\hat\beta_1$} &  \multicolumn{2}{c}{$\hat\alpha_0$} & \multicolumn{2}{c}{$\hat\alpha_1$} & \multicolumn{2}{c}{$\hat\nu_0$} \\[-0.0cm]
 \hline
&&&&&&&&&&&&&&\\[-0.3cm]
   0.00  &   0.00 &   ---  &   0.00 &   ---  &   0.88 & (0.16) &   0.03 & (0.12) &   0.17 & (0.08) &   -0.51 & (0.08) &   -0.35 & (0.18)  \\[-0.0cm]
   0.22  &   0.00 &   ---  &   0.00 &   ---  &   0.91 & (0.17) &   0.04 & (0.11) &   0.19 & (0.08) &   -0.51 & (0.08) &   -0.07 & (0.23)  \\[-0.0cm]
   0.41  &   0.00 &   ---  &   0.00 &   ---  &   0.94 & (0.18) &   0.04 & (0.12) &   0.19 & (0.08) &   -0.51 & (0.08) &    0.22 & (0.30)  \\[-0.0cm]
   0.69  &   0.00 &   ---  &   0.00 &   ---  &   0.98 & (0.20) &   0.07 & (0.14) &   0.20 & (0.09) &   -0.50 & (0.09) &    0.72 & (0.94)  \\[-0.0cm]
   1.10  &   0.00 &   ---  &   0.00 &   ---  &   1.04 & (0.17) &   0.08 & (0.15) &   0.22 & (0.08) &   -0.50 & (0.09) &    1.74 & (5.10)  \\[-0.0cm]
   1.61  &   0.00 &   ---  &   0.00 &   ---  &   1.20 & (0.15) &   0.09 & (0.17) &   0.27 & (0.07) &   -0.49 & (0.10) &   14.00 & (7.03)  \\[-0.0cm]
$\infty$ &   0.00 &   ---  &   0.00 &   ---  &   1.54 & (0.15) &   0.12 & (0.20) &   0.37 & (0.08) &   -0.49 & (0.10) &   16.85 & (1.76)  \\[0.1cm]
 \hline
&&&&&&&&&&&&&&\\[-0.2cm]
\multicolumn{15}{p{.95\textwidth}}{\footnotesize All numbers are rounded to two decimal places. For the models with fixed parameters, the ``estimated'' value shown is the value at which the parameter is fixed, and its standard error is then indicated by ``---''.}
\end{tabular}
\end{footnotesize}
\end{center}
\end{table}

\begin{table}[htbp]
\begin{center}
\caption{Median and standard {error} (in brackets) of estimates when $n=100$}
\begin{footnotesize}
\begin{tabular}{c@{\qquad}r@{~}c@{~~}r@{~}c@{~~}r@{~}c@{~~}r@{~}c@{~~}r@{~}c@{~~}r@{~}c@{~~}r@{~}c}
\hline
&&&&&&&&&&&&&&\\[-0.3cm]
& \multicolumn{14}{c}{\underline{Model$(\tau, \beta, \alpha)$}} \\[-0.0cm]
&&&&&&&&&&&&&&\\[-0.3cm]
$\nu$ &  \multicolumn{2}{c}{$\hat\tau_0$} & \multicolumn{2}{c}{$\hat\tau_1$} & \multicolumn{2}{c}{$\hat\beta_0$} & \multicolumn{2}{c}{$\hat\beta_1$} & \multicolumn{2}{c}{$\hat\alpha_0$} &  \multicolumn{2}{c}{$\hat\alpha_1$} & \multicolumn{2}{c}{$\hat\nu_0$} \\[-0.0cm]
 \hline
&&&&&&&&&&&&&&\\[-0.3cm]
   0.00  &   1.32 & (2.72) &   0.88 & (2.91) &   -0.49 & (3.18) &   -0.01 & (1.97) &   0.39 & (0.43) &   -0.52 & (0.63) &    0.08 & (2.53) \\[-0.0cm]
   0.22  &   1.29 & (2.21) &   0.85 & (3.20) &   -0.39 & (2.74) &    0.01 & (2.39) &   0.38 & (0.42) &   -0.49 & (0.56) &    0.17 & (3.18) \\[-0.0cm]
   0.41  &   1.10 & (1.85) &   0.83 & (3.21) &   -0.22 & (2.24) &    0.02 & (2.73) &   0.35 & (0.44) &   -0.51 & (0.52) &    0.24 & (3.41) \\[-0.0cm]
   0.69  &   0.87 & (1.94) &   0.67 & (3.47) &   -0.14 & (2.48) &   -0.03 & (3.12) &   0.21 & (0.45) &   -0.49 & (0.46) &    0.87 & (6.34) \\[-0.0cm]
   1.10  &   0.61 & (1.52) &   0.50 & (2.84) &    0.05 & (1.96) &   -0.02 & (2.69) &   0.17 & (0.33) &   -0.52 & (0.37) &   12.66 & (7.62) \\[-0.0cm]
   1.61  &   0.70 & (1.22) &   0.57 & (2.46) &    0.05 & (1.72) &    0.05 & (2.52) &   0.17 & (0.28) &   -0.52 & (0.37) &   15.31 & (7.02) \\[-0.0cm]
$\infty$ &   0.89 & (0.89) &   0.75 & (1.81) &   -0.16 & (1.44) &   -0.14 & (2.05) &   0.23 & (0.26) &   -0.53 & (0.34) &   15.95 & (6.12) \\[0.1cm]
\hline
&&&&&&&&&&&&&&\\[-0.3cm]
& \multicolumn{14}{c}{\underline{Model$(\tau, \alpha)$}}   \\[-0.0cm]
&&&&&&&&&&&&&&\\[-0.3cm]
$\nu$ &  \multicolumn{2}{c}{$\hat\tau_0$} & \multicolumn{2}{c}{$\hat\tau_1$} & \multicolumn{2}{c}{$\hat\beta_0$} & \multicolumn{2}{c}{$\hat\beta_1$} & \multicolumn{2}{c}{$\hat\alpha_0$} &  \multicolumn{2}{c}{$\hat\alpha_1$} & \multicolumn{2}{c}{$\hat\nu_0$} \\[-0.0cm]
 \hline
&&&&&&&&&&&&&&\\[-0.3cm]
   0.00  &   0.79 & (0.29) &   0.59 & (0.44) &   0.00 &   ---  &   0.00 &   ---  &   0.24 & (0.20) &   -0.50 & (0.20) &   -0.02 & (0.27) \\[-0.0cm]
   0.22  &   0.79 & (0.28) &   0.58 & (0.39) &   0.00 &   ---  &   0.00 &   ---  &   0.21 & (0.20) &   -0.50 & (0.21) &    0.22 & (1.47) \\[-0.0cm]
   0.41  &   0.76 & (0.27) &   0.58 & (0.35) &   0.00 &   ---  &   0.00 &   ---  &   0.21 & (0.20) &   -0.51 & (0.20) &    0.45 & (1.93) \\[-0.0cm]
   0.69  &   0.79 & (0.26) &   0.58 & (0.31) &   0.00 &   ---  &   0.00 &   ---  &   0.22 & (0.20) &   -0.51 & (0.20) &    0.72 & (3.68) \\[-0.0cm]
   1.10  &   0.82 & (0.24) &   0.59 & (0.25) &   0.00 &   ---  &   0.00 &   ---  &   0.24 & (0.20) &   -0.51 & (0.20) &    1.07 & (5.51) \\[-0.0cm]
   1.61  &   0.83 & (0.20) &   0.63 & (0.24) &   0.00 &   ---  &   0.00 &   ---  &   0.26 & (0.19) &   -0.52 & (0.19) &    1.54 & (6.86) \\[-0.0cm]
$\infty$ &   0.89 & (0.16) &   0.66 & (0.19) &   0.00 &   ---  &   0.00 &   ---  &   0.30 & (0.18) &   -0.50 & (0.21) &    4.00 & (7.52) \\[0.1cm]
\hline
&&&&&&&&&&&&&&\\[-0.3cm]
& \multicolumn{14}{c}{\underline{Model$(\beta, \alpha)$}}   \\[-0.0cm]
&&&&&&&&&&&&&&\\[-0.3cm]
$\nu$ &  \multicolumn{2}{c}{$\hat\tau_0$} & \multicolumn{2}{c}{$\hat\tau_1$} & \multicolumn{2}{c}{$\hat\beta_0$} & \multicolumn{2}{c}{$\hat\beta_1$} &  \multicolumn{2}{c}{$\hat\alpha_0$} & \multicolumn{2}{c}{$\hat\alpha_1$} & \multicolumn{2}{c}{$\hat\nu_0$} \\[-0.0cm]
 \hline
&&&&&&&&&&&&&&\\[-0.3cm]
   0.00  &   0.00 &   ---  &   0.00 &   ---  &   0.89 & (0.36) &   0.04 & (0.25) &   0.19 & (0.17) &   -0.51 & (0.17) &   -0.34 & (0.44)  \\[-0.0cm]
   0.22  &   0.00 &   ---  &   0.00 &   ---  &   0.89 & (0.38) &   0.07 & (0.28) &   0.19 & (0.18) &   -0.50 & (0.18) &    0.01 & (0.79)  \\[-0.0cm]
   0.41  &   0.00 &   ---  &   0.00 &   ---  &   0.92 & (0.38) &   0.03 & (0.30) &   0.19 & (0.19) &   -0.51 & (0.19) &    0.24 & (1.99)  \\[-0.0cm]
   0.69  &   0.00 &   ---  &   0.00 &   ---  &   1.04 & (0.38) &   0.06 & (0.33) &   0.23 & (0.17) &   -0.51 & (0.19) &    0.70 & (3.81)  \\[-0.0cm]
   1.10  &   0.00 &   ---  &   0.00 &   ---  &   1.21 & (0.35) &   0.07 & (0.33) &   0.29 & (0.16) &   -0.50 & (0.20) &    1.27 & (5.92)  \\[-0.0cm]
   1.61  &   0.00 &   ---  &   0.00 &   ---  &   1.35 & (0.36) &   0.09 & (0.38) &   0.34 & (0.17) &   -0.50 & (0.21) &    2.32 & (7.23)  \\[-0.0cm]
$\infty$ &   0.00 &   ---  &   0.00 &   ---  &   1.66 & (0.36) &   0.13 & (0.43) &   0.42 & (0.17) &   -0.50 & (0.22) &   15.98 & (6.32)  \\[0.1cm]
 \hline
 &&&&&&&&&&&&&&\\[-0.2cm]
\multicolumn{15}{p{.95\textwidth}}{\footnotesize All numbers are rounded to two decimal places. For the models with fixed parameters, the ``estimated'' value shown is the value at which the parameter is fixed, and its standard error is then indicated by ``---''.}
\end{tabular}
\end{footnotesize}
\end{center}
\end{table}

\end{document}